\documentclass[12pt]{article}
\usepackage{epsfig}
\usepackage{psfrag}
\usepackage{enumitem}
\usepackage{latexsym}
\usepackage{indentfirst}
\usepackage{fancyhdr}
\usepackage{amssymb}
\usepackage{amsmath}
\usepackage{amsfonts}
\usepackage{pifont}
\usepackage{cite}
\usepackage{bbold}
\usepackage{color}
\usepackage[footnotesize]{caption2}
\usepackage{graphicx}
\usepackage[center,footnotesize,hang]{subfigure}
\usepackage{url}
\usepackage{array}

\begin{document}
\begin{titlepage}
\vspace*{-1cm}

\hfill{MADPH-11-1576}

\vskip 2.5cm
\begin{center}
{\Large\bf Golden Ratio Neutrino Mixing and $A_5$ Flavor Symmetry}
\end{center}
\vskip 0.2  cm
\vskip 0.5  cm
\begin{center}
{ Gui-Jun Ding}$^{a,b}$,
{Lisa L. Everett}$^{b}$, {and Alexander J. Stuart}$^{b,c}$
\\
\vskip .2cm
$^{a}${\it Department of Modern Physics,}
\\
{\it University of Science and Technology of China, Hefei, Anhui
230026, China}
 \vskip .2cm
$^b${\it Department of Physics, University of
Wisconsin-Madison,}\\
{\it 1150 University Avenue, Madison, WI 53706, USA}\vskip.2cm
$^c${\it School of Physics and Astronomy, University of Southampton, Southampton, SO17 1BJ, United Kingdom}\vskip.2cm

\end{center}
\vskip 0.7cm
\begin{abstract}
\noindent

We provide a systematic and thorough exploration of lepton flavor models in which the solar mixing angle is related to the golden ratio.  For scenarios in which the solar mixing angle is given by the inverse cotangent of the golden ratio, we demonstrate that $A_5$ is the smallest non-Abelian finite group that contains all of the symmetries necessary to enforce this specific lepton mixing pattern.  Within this context, we propose two lepton flavor models that yield this mixing pattern through the breaking of $A_5$ at leading order to the Klein four subgroup in the neutrino sector.  Both models have triplet embeddings of the lepton doublets as well as the charged lepton singlets.  In the charged lepton sector, the residual symmetry is $Z_5$ in the first model, while in the second model, $A_5$ is broken completely at leading order.  For the second model, the reactor mixing angle vanishes at leading order and is of the order of the square of the Cabibbo angle at next-to-leading order, which is allowed by the global analysis of current lepton data.

\end{abstract}
\end{titlepage}
\setcounter{footnote}{0}
\vskip2truecm

\section{Introduction}
Neutrino detection experiments have now provided definitive evidence that the solar and atmospheric neutrino anomalies can be explained by neutrino oscillations. Global fits of the data for the mass-squared differences and the mixing angles of the Maki-Nakagawa-Sakata-Pontecorvo (MNSP) matrix \cite{Maki:1962mu,Pontecorvo:1967fh} assuming three families of neutrinos have been provided by several groups \cite{Schwetz:2011zk,Schwetz:2011qt,Fogli:2011qn,Fogli:Indication,GonzalezGarcia:2010er}; we show the results of the latest two independent global fits in Table~\ref{tab:summary2011}.
\begin{table}[hptb]\centering
\begin{tabular}{|c|cc|cc|} \hline\hline
 & \multicolumn{2}{|c|}{Ref.~\cite{Schwetz:2011zk}}
        & \multicolumn{2}{|c|}{Ref.~\cite{Fogli:2011qn}}\\
parameter   & best fit$\pm 1\sigma$ & 3$\sigma$ interval   & best
fit$\pm 1\sigma$ & 3$\sigma$ interval   \\ \hline

  $\Delta m^2_{21}\: [10^{-5}{\rm eV^2}]$ & $7.59^{+0.20}_{-0.18}$ & 7.09 -- 8.19
    & $7.58^{+0.22}_{-0.26}$ & 6.99 -- 8.18 \\[1.5mm]

  $|\Delta m^2_{31}|\: [10^{-3}{\rm eV^2}]$
        &
        \begin{tabular}{c}
        $2.50^{+0.09}_{-0.16}$\\
        $2.40^{+0.08}_{-0.09}$
        \end{tabular} & \hspace{-9pt}
        \begin{tabular}{c}
        2.14 -- 2.76\\
        2.13 -- 2.67
        \end{tabular}
    &  $2.35^{+0.12}_{-0.09}$
    & \hspace{-11pt}  2.06 -- 2.67
    \\[3mm]

 $\sin^2\theta_{12}$
        & $0.312^{+0.017}_{-0.015}$ & 0.27 -- 0.36
        &
        \begin{tabular}{c}
        $0.306^{+0.018}_{-0.015}$\\
        $(0.312^{+0.017}_{-0.016})$
        \end{tabular} &
        \begin{tabular}{c}
        0.259 -- 0.359\\
        (0.265 -- 0.364)
        \end{tabular}
    \\[4mm]

 $\sin^2\theta_{23}$
        &
        \begin{tabular}{c}
        $0.52^{+0.06}_{-0.07}$\\
        $0.52\pm0.06$
        \end{tabular}
         & 0.39 -- 0.64
    & $0.42^{+0.08}_{-0.03}$ & 0.34 -- 0.64
    \\[4mm]

 $\sin^2\theta_{13}$
        &
        \begin{tabular}{c}
        $0.013^{+0.007}_{-0.005}$\\
        $0.016^{+0.008}_{-0.006}$
        \end{tabular}
        &
        \begin{tabular}{c}
        0.001 -- 0.035\\
        0.001 -- 0.039
        \end{tabular}
        &
        \begin{tabular}{c}
        $0.021^{+0.007}_{-0.008}$\\
        $(0.025\pm0.007)$
        \end{tabular}
         &
         \begin{tabular}{c}
         0.001 -- 0.044 \\
         (0.005 -- 0.050)
         \end{tabular}
    \\
        \hline\hline

\end{tabular}
\caption{\label{tab:summary2011} Summary of neutrino oscillation parameters from two global data fits \cite{Schwetz:2011zk,Fogli:2011qn}.
The upper (lower) row corresponds to normal (inverted) neutrino mass hierarchy in \cite{Schwetz:2011zk}, and the results from the new reactor fluxes are shown in parentheses for the fit of \cite{Fogli:2011qn}.}
\end{table}

The neutrino oscillation data have revealed great differences between the quark and lepton mixing angle patterns. The atmospheric angle $\theta_{23}$ is compatible with maximal mixing, but the error bars admit relatively large deviations. The solar angle $\theta_{12}$ is more precisely measured and is known to be large but not maximal at the $5\sigma$ level. The reactor angle $\theta_{13}$ has recently been the source of great excitement since T2K \cite{Abe:2011sj} and MINOS \cite{minos} each reported evidence for a relatively large $\theta_{13}$ at the level of $2.5\sigma$ and $1.7\sigma$, respectively. The first result from Double Chooz experiment also implies a non-zero $\theta_{13}$ \cite{double_chooz}. The global fits quoted above that include these new results indicate $\theta_{13}>0$ at about the $3\sigma$ level, but the magnitude of $\theta_{13}$ still suffers from large uncertainties.  Future measurements of $\theta_{13}$ at Double Chooz \cite{Ardellier:2006mn} and Daya Bay \cite{Wang:2006ca} will provide important clues in the search for a greater understanding of the fermion mass puzzle of the Standard Model (SM).

In contrast to the small quark mixing angles, the large lepton mixing angles suggest a model-building paradigm based on discrete non-Abelian family symmetries (for reviews, see \cite{review}).  Lepton models  have been constructed based on the tetrahedral group $T$ (which is isomorphic to $A_4$) \cite{A4},  the binary tetrahedral group $T^\prime$ \cite{tprime},  $\Delta(3n^2)$ and $\Delta(6n^2)$ \cite{delta}, the semidirect product of $Z_3$ and $Z_7$ \cite{Z3xZ7}, $Z_2\times Z_2$ \cite{Kajiyama:2007gx}, $S_4$ \cite{S4},  $S_3$ \cite{S3}, the semidirect product of $A_4$ and $S_3$ \cite{S3xA4}, $PSL(2,7)$ \cite{PSL27},  the quaternionic symmetries \cite{Quat}, the dihedral symmetries $D_n$ \cite{D,GR2},  $T_{13}$ \cite{T13}, and the icosahedral group $I$ (which is isomorphic to $A_5$)  \cite{A5everettstuart,A54fam,Feruglio:2011qq} (an extension to $I^\prime$ for the quark sector can be found in \cite{Everett:2010rd}). Many scenarios lead to the well-known Harrison-Perkins-Scott [HPS] ``tri-bimaximal" mixing pattern  \cite{TBmix}, in which at leading order $\theta_{23}$ is maximal, $\theta_{13}$ is zero, and the solar angle is given by $\cot\theta_{12}=\sqrt{2}$.  We note that the tri-bimaximal matrix appeared earlier in the context of quark masses \cite{TB_quark}.
However, there are also alternative mixing patterns that are also in agreement with current data, but differ according to their leading order predictions for the solar mixing angle $\theta_{12}$.  Recently, a number of intriguing scenarios have also been proposed in which $\theta_{13}\neq 0 $ at leading order \cite{S4t13,A4t13,S3t13,SU5t13,A4S4t13,model.ind.t13,Dt13}, in response to the recent possible hints of nonvanishing $\theta_{13}$.

In this paper, we focus on scenarios in which at leading order $\theta_{23}$ is maximal, $\theta_{13}=0$, and $\theta_{12}$  is governed by the golden ratio, $\phi_g=(1+\sqrt{5})/2$.  In this context, there are two ideas that have been proposed. The first, which we call ``GR1" mixing, is the hypothesis that
\begin{equation}
\cot\theta_{12}=\phi_g,
\end{equation}
which was first proposed in  \cite{GRPrediction1} and later explored in detail within a $Z_2\times Z_2$ flavor symmetry in \cite{Kajiyama:2007gx}.  In this case, $\sin^2\theta_{12}=\sqrt{5}/(5\phi_g)\simeq0.276$,
which is  very close to the $3\sigma$ lower bound.\footnote{$\sin^2\theta_{12}\simeq0.276$ is slightly above the $3\sigma$ lower bound of the first global fit, while it is around the $2\sigma$ lower limit of the second fit.}  As first suggested in \cite{Kajiyama:2007gx},  the icosahedral symmetry group $\mathcal{I}\sim A_5$ is a natural candidate as a flavor symmetry that can yield this mixing pattern, since the golden ratio
characterizes several geometrical properties of the icosahedron.  As a result, $A_5$ models based on this idea have been proposed in \cite{A5everettstuart,A54fam,Feruglio:2011qq}.  The second, which we call ``GR2" mixing, is the relation first suggested by \cite{Rodejohann:2008ir}, in which
\begin{equation}
\cos\theta_{12}=\frac{\phi_g}{2}.
\end{equation}
Hence, in this case $\sin^2\theta_{12}= \sqrt{5}/(4\phi_g)\simeq0.345$, which is close to the $2\sigma$ upper limit. A model realization
of this mixing pattern based on dihedral symmetry groups was given in \cite{GR2}.

The purpose of our work is to provide a systematic exploration of the golden ratio mixing hypotheses.  For both of the GR1 and GR2 mixing patterns,  we examine the question of what is the minimal group that naturally contains all of the symmetries of the lepton matrices within each scenario.  In the case of GR1 mixing, we find after a systematic analysis that the minimal group is indeed $A_5$.   In the case of GR2 mixing, there is no discrete group (up to a very large order) that contains the complete set of symmetries of the associated lepton mass matrices.  Focusing our attention on the case of GR1 mixing, we then provide a comprehensive exploration of lepton flavor model building based on $A_5$, with the key input that the lepton doublets and charged lepton singlets of the SM are embedded in {\it triplet} representations of $A_5$ as opposed to singlet embeddings.  Within the framework of supersymmetric models, we obtain two models of this type that differ in terms of their respective residual symmetries in the charged lepton sector.  In one model, $A_5$ is broken to $Z_5$ in the charged lepton sector and to Klein four in the neutrino sector, resulting naturally in GR1 mixing.  However, fine-tuning is needed to obtain the charged lepton mass hierarchy.  The second model improves on this scenario by breaking this residual symmetry.

This paper is organized as follows. In Section 2, we present our investigation of the minimal family symmetry group required to produce the GR mixing patterns.  As this procedure yields the minimal group of $A_5$ for GR1 mixing and no obvious solution for GR2 mixing, in
Section 3 we present basic group theoretic aspects of $A_5$ that are useful for flavor model-building, including the equivalency classes and the explicit subgroups. In Section 4, we turn to $A_5$ flavor model-building in the lepton sector. We summarize and conclude in Section 5.   In the Appendix, further practical details of the group theory of $A_5$ are given, including the explicit representation matrices and the Clebsch-Gordan coefficients.

\section{\label{sec:minimal}Minimal Flavor Symmetry for Golden Ratio Mixing}
In this section, we address the question of finding the minimal discrete non-Abelian flavor symmetry group that can naturally realize the golden ratio mixing hypotheses.  In this context, the minimal group is defined as the smallest finite group that contains as elements all of the symmetries of the neutral and charged lepton mass matrices.   Given these symmetries, we then solve the inverse of the usual issue of finding group representations:  given specific representation matrices, it is necessary determine the group such that these matrices form a faithful representation of the group ({\it i.e.}, one in which the kernel of the representation is just the trivial subgroup consisting of the group's identity element).

The first step is to determine the symmetries of the mass matrices. We start with the charged lepton sector.  We choose to work in a basis which $\widetilde{m}_{\ell}\equiv m^{\dagger}_{\ell}m_{\ell}$ is diagonal, where $m_{\ell}$ is the charged lepton mass matrix. Hence, it is straightforward to see that the symmetry transformation matrix $G_{\ell}$, which is determined by the condition $G^{\dagger}_{\ell}\widetilde{m}_{\ell}G_{\ell}=\widetilde{m}_{\ell}$, is a diagonal and non-degenerate $3\times3$ unitary matrix ({\it i.e.}, a diagonal and non-degenerate phase matrix).  This ensures that $\widetilde{m}_{\ell}$ remains diagonal since it has non-degenerate eigenvalues.

Turning to the neutrino sector, with this choice of basis the neutrino mass matrix can be reconstructed in terms of its complex mass eigenvalues $m_{1,2,3}$ as follows:
\begin{equation}
m_{\nu}=U^{*}{\rm diag}(m_1,m_2,m_3)U^{\dagger}.
\end{equation}
Here $U$ is the MNSP matrix (in a specific phase convention) with $\theta_{12}$ unspecified, $\theta_{23}=\pi/4$, and $\theta_{13}=0$:
\begin{equation}
U=\left(\begin{array}{ccc} c_{12} & s_{12}  &0\\
-s_{12}/\sqrt{2}  & c_{12}/\sqrt{2} & 1/\sqrt{2}\\
-s_{12}/\sqrt{2}  & c_{12}/\sqrt{2} & -1/\sqrt{2}
\end{array}\right),
\label{mnspgen}
\end{equation}
in which $s_{12}=\sin\theta_{12}$ and $c_{12}=\cos\theta_{12}$.\footnote{We
absorb the Majorana phases in the mass eigenvalues $m_i$, rather
than in the MNSP matrix.}  The neutrino mass matrix then takes the form
\begin{eqnarray}
m_{\nu}=\left(\begin{array}{ccc} m_1c^2_{12}+m_2s^2_{12} &
-\frac{m_1-m_2}{\sqrt{2}}s_{12}c_{12} &
-\frac{m_1-m_2}{\sqrt{2}}s_{12}c_{12}\\
-\frac{m_1-m_2}{\sqrt{2}}s_{12}c_{12} & \frac{1}{2}m_1s^2_{12}+\frac{1}{2}m_2c^2_{12}+\frac{1}{2}m_3& \frac{1}{2}m_1s^2_{12}+\frac{1}{2}m_2c^2_{12}-\frac{1}{2}m_3\\
-\frac{m_1-m_2}{\sqrt{2}}s_{12}c_{12} &
\frac{1}{2}m_1s^2_{12}+\frac{1}{2}m_2c^2_{12}-\frac{1}{2}m_3 &
\frac{1}{2}m_1s^2_{12}+\frac{1}{2}m_2c^2_{12}+\frac{1}{2}m_3
\end{array}\right).
\end{eqnarray}

We turn now to the GR1 mixing pattern, in which  $\cot\theta_{12}=\phi_g$ (a similar analysis can be found in \cite{Feruglio:2011qq}).  In this case, the neutrino mass matrix is given by
\begin{eqnarray}
m^{{\rm GR}1}_{\nu}&=&\frac{m_1}{2\sqrt{5}}\left(\begin{array}{ccc}
2\phi_g& -\sqrt{2}  & -\sqrt{2} \\
-\sqrt{2}  &  -1/\phi_g &  -1/\phi_g\\
-\sqrt{2}  & -1/\phi_g  & -1/\phi_g
\end{array}\right)
+\frac{m_2}{2\sqrt{5}}\left(\begin{array}{ccc}
2/\phi_g & \sqrt{2}   & \sqrt{2}\\
\sqrt{2}   & \phi_g   & \phi_g \\
\sqrt{2}   & \phi_g   & \phi_g
\end{array}\right)\nonumber \\
\label{5}
&+&\frac{m_3}{2}\left(\begin{array}{ccc} 0&0&0\\
0&\;1&\!\!-1\\
0&\!\!-1&\;1
\end{array}\right).
\end{eqnarray}
The $2-3$ symmetry of Eq.~(\ref{5}) is due to the maximal atmospheric angle and the vanishing reactor angle. In addition, the
following relation is satisfied:
\begin{equation}
\label{6}(m^{{\rm GR}1}_{\nu})_{11}+\sqrt{2}\,(m^{{\rm GR}1}_{\nu})_{12}=(m^{{\rm GR}1}_{\nu})_{22}+(m^{{\rm GR}1}_{\nu})_{23}.
\end{equation}
To obtain the symmetry properties of $m^{{\rm GR}1}_{\nu}$, one seeks the unitary transformations $G_i$ that satisfy the condition
$G^{T}_{i}m^{{\rm GR}1}_{\nu}G_{i}=m^{{\rm GR}1}_{\nu}$.  After a straightforward computation, we find (in addition to the identity transformation) three unitary transformation matrices, $G_{1,2,3}$:\footnote{We have chosen a convention for the overall sign of each $G_i$ such that its determinant is equal to unity.}
\begin{eqnarray}
\nonumber&&
G_1=\frac{1}{\sqrt{5}}\left(\begin{array}{ccc} 1  &
-\sqrt{2} &
-\sqrt{2} \\
-\sqrt{2}  & -\phi_g  &  1/\phi_g\\
-\sqrt{2}  & 1/\phi_g  & -\phi_g
\end{array}\right),\;\;
G_2=\frac{1}{\sqrt{5}}\left(\begin{array}{ccc}
-1 & \sqrt{2}  & \sqrt{2}\\
\sqrt{2}  & -1/\phi_g  & \phi_g\\
\sqrt{2}  & \phi_g  & -1/\phi_g
\end{array}\right)\\
\label{7}&&
G_3=-\left(\begin{array}{ccc} 1&0&0\\
0&0&1\\
0&1&0
\end{array}\right),
\end{eqnarray}
which satisfy the relations
\begin{equation}
\label{k4} G^2_{i}=\mathbf{1},~~~~~~~G_iG_j=G_jG_i=G_{k}~~{\rm
with}~~ i\neq j\neq k,
\end{equation}
Therefore, the symmetry group of the neutrino mass matrix $m^{{\rm GR}1}_{\nu}$
is the Klein four group $K_4\simeq Z_2\times Z_2$ (the result that the symmetry group is $Z_2\times Z_2$  was also found in \cite{Feruglio:2011qq}).

For the GR2 mixing pattern with $\cos\theta_{12}=\phi_g/2$, an equivalent reconstruction of the neutrino mass matrix results in
\begin{eqnarray}
&&m^{{\rm GR}2}_{\nu}=\frac{m_1}{8}\left(\begin{array}{ccc}
2(1+\phi_g) &-\sqrt{4+2\phi_g} &  -\sqrt{4+2\phi_g}\\
-\sqrt{4+2\phi_g} & 3-\phi_g & 3-\phi_g \\
-\sqrt{4+2\phi_g} &  3-\phi_g &  3-\phi_g\end{array}\right)\nonumber \\ &&+
\frac{m_2}{8}\left(\begin{array}{ccc} 2(3-\phi_g) &
\sqrt{4+2\phi_g} &  \sqrt{4+2\phi_g}\\
\sqrt{4+2\phi_g} & 1+\phi_g  & 1+\phi_g\\
\sqrt{4+2\phi_g} & 1+\phi_g  & 1+\phi_g
\end{array}\right)
+\frac{m_3}{2}\left(\begin{array}{ccc} 0&0&0 \\
0&\;1&\!\!-1\\
0&\!\!-1&\;1
\end{array}\right).
\end{eqnarray}
This matrix, which is also clearly $2-3$ symmetric, satisfies the following sum rule:
\begin{equation}
\label{add2}
(m_{\nu}^{GR2})_{11}+2\sqrt{\frac{2}{\phi_g^3\sqrt{5}}}(m_{\nu}^{GR2})_{12}
=(m^{{\rm GR}2}_{\nu})_{22}+(m^{{\rm GR}2}_{\nu})_{23}.
\end{equation}
Following the same procedure as above for determining the symmetry transformations $G_i'$ that satisfy $G'^{T}_{i}m^{{\rm GR}2}_{\nu}G'_{i}=m^{{\rm GR}2}_{\nu}$, we find the following set of three unitary matrices:
\begin{eqnarray}\
\nonumber&&
G'_1=\frac{1}{2}\left(\begin{array}{ccc} 1/\phi_g &
-\sqrt{\frac{\sqrt{5}}{2} \phi_g}  &  -\sqrt{\frac{\sqrt{5}}{2} \phi_g}  \\
-\sqrt{\frac{\sqrt{5}}{2} \phi_g}   & -\phi_g^2 & \frac{\sqrt{5}}{2} \frac{1}{\phi_g} \\
-\sqrt{\frac{\sqrt{5}}{2} \phi_g}  &  \frac{\sqrt{5}}{2} \frac{1}{\phi_g} & -\phi_g^2
\end{array}\right),\;
G'_2=\frac{1}{2}\left(\begin{array}{ccc}-1/\phi_g &
\sqrt{\frac{\sqrt{5}}{2} \phi_g} & \sqrt{\frac{\sqrt{5}}{2} \phi_g}  \\
\sqrt{\frac{\sqrt{5}}{2} \phi_g}  & -\frac{\sqrt{5}}{2} \frac{1}{\phi_g}   & \phi_g^2  \\
\sqrt{\frac{\sqrt{5}}{2} \phi_g}  & \phi_g^2 &  -\frac{\sqrt{5}}{2} \frac{1}{\phi_g}
\end{array}\right)\\
\label{11}&&G'_3=-\left(\begin{array}{ccc}1&0&0\\
0&0&1\\
0&1&0
\end{array}\right).
\end{eqnarray}
The $G'_i$ satisfy the relations given in Eq.~(\ref{k4}), so the symmetry of the neutrino mass matrix $m^{{\rm GR}2}_{\nu}$ is also the Klein four group.  Note also that $G'_3=G_3$; this is as expected since it is determined solely by the requirements that $\theta_{13}=0$ and $\theta_{23}=\pi/4$.

Based on these results, we apply the following procedure to find the minimal discrete family symmetry group for these mixing patterns.  Denoting this underlying family symmetry group at high energies as ${\cal G}$, the above considerations tell us that $G_\ell$ and $G_{1,2,3}$ ($G'_{1,2,3}$) should be elements of this group for GR1 (GR2) mixing in order that they are preserved in the theory at lower scales at leading order (LO).  For the case of  finite groups, this implies that there must be some integers $n$ and $m_i$ such that
\begin{equation}
\label{subgroup}
G^{n}_{\ell}=(G_iG_{\ell})^{m_i}=\mathbf{1},\qquad n\geq3,
\end{equation}
where the restriction on $n$ results from the requirement that $G_\ell$ is nondegenerate, and an identical relation holds for the primed version $G'_{1,2,3}$.  The group generated in this way for each $G_i$ must be a subgroup of the full symmetry group ${\cal G}$.  Hence, the procedure is first to determine possible solutions for $G_\ell$ that result in the relation of Eq.~(\ref{subgroup}) with integer values for $n$ and $m_i$ given $G_i$  or $G'_i$, and then to examine their possible embeddings in discrete groups.   Working in the diagonal basis for the charged leptons leads to no loss of generality because the results obtained in different bases are related via similarity transformations.  Furthermore, the $2-3$ symmetry of the neutrino mass matrices  indicates that identical results are obtained for choices of $G_\ell$ with the (22) and (33) elements of $G_\ell$ interchanged.

Let us consider the case of GR1 mixing as the first example.  An explicit computation yields the result that for $n=3$ ($G^{3}_{\ell}=\mathbf{1}$) or $n=4$  ($G^{4}_{\ell}=\mathbf{1}$), the resulting family symmetry ${\cal G}$ must be an infinite group, since there is no integer $m$ for which $(G_1G_{\ell})^{m}=\mathbf{1}$ or $(G_2G_{\ell})^{m}=\mathbf{1}$.  However, $n=5$ ($G^{5}_{\ell}=\mathbf{1}$) yields ten nontrivial solutions, which are tabulated below:

\begin{itemize}
\item $G_{\ell}=(1,\rho,\rho^4)$:   $(G_1G_\ell)^3=\mathbf{1}$, $(G_2G_\ell)^5=\mathbf{1}$, $(G_3G_\ell)^2=\mathbf{1}$.

\item $G_{\ell}=(1,\rho^2,\rho^3)$:   $(G_1G_\ell)^5=\mathbf{1}$, $(G_2G_\ell)^3=\mathbf{1}$, $(G_3G_\ell)^2=\mathbf{1}$.

\item $G_{\ell}=(\rho,1,\rho^2)$:   $(G_1G_\ell)^{15}=\mathbf{1}$, $(G_2G_\ell)^5=\mathbf{1}$, $(G_3G_\ell)^{10}=\mathbf{1}$.
\item $G_{\ell}=(\rho^4,1,\rho^3)$:   $(G_1G_\ell)^{15}=\mathbf{1}$, $(G_2G_\ell)^5=\mathbf{1}$, $(G_3G_\ell)^{10}=\mathbf{1}$.
\item $G_{\ell}=(\rho^2,\rho,\rho^3)$:   $(G_1G_\ell)^{15}=\mathbf{1}$, $(G_2G_\ell)^5=\mathbf{1}$, $(G_3G_\ell)^{10}=\mathbf{1}$.
\item $G_{\ell}=(\rho^3,\rho^2,\rho^4)$:   $(G_1G_\ell)^{15}=\mathbf{1}$, $(G_2G_\ell)^5=\mathbf{1}$, $(G_3G_\ell)^{10}=\mathbf{1}$.

\item $G_{\ell}=(\rho^2,1,\rho^4)$:   $(G_1G_\ell)^{5}=\mathbf{1}$, $(G_2G_\ell)^{15}=\mathbf{1}$, $(G_3G_\ell)^{10}=\mathbf{1}$.
\item $G_{\ell}=(\rho^3, 1,\rho)$:   $(G_1G_\ell)^{5}=\mathbf{1}$, $(G_2G_\ell)^{15}=\mathbf{1}$, $(G_3G_\ell)^{10}=\mathbf{1}$.
\item $G_{\ell}=(\rho, \rho^3,\rho^4)$:   $(G_1G_\ell)^{5}=\mathbf{1}$, $(G_2G_\ell)^{15}=\mathbf{1}$, $(G_3G_\ell)^{10}=\mathbf{1}$.
\item $G_{\ell}=(\rho^4, \rho,\rho^2)$:   $(G_1G_\ell)^{5}=\mathbf{1}$, $(G_2G_\ell)^{15}=\mathbf{1}$, $(G_3G_\ell)^{10}=\mathbf{1}$.

\end{itemize}
Here $\rho=e^{2\pi i/5}$, and we write $G_\ell= \mathrm{diag}(a,b,c)$ by $(a,b,c)$ for notational simplicity.  There are two sets of solutions:  one with $m_i=(5,3,2)$ (or $(3,5,2)$), and one with $m_i=(15,5,10)$ (or $(5,15,10)$).  Within each set, the two permutations of the $m_i$ are connected because the corresponding $G_\ell$'s are related.  For example,  for the first set of solutions the two options for $G_\ell$, $G_{\ell_1}={\rm diag}(1,\rho,\rho^4)$ and $G_{\ell_2}={\rm diag}(1, \rho^2,\rho^3)$, satisfy $G_{\ell_1}=G^3_{\ell_2}$ and $G_{\ell_2}=G^{2}_{\ell_1}$.  Hence, they must generate the same group structure.  Similar relations exist for the second set of solutions; in addition, the $G_\ell$'s can be obtained by multiplying either $G_{\ell_1}$ or $G_{\ell_2}$ by $\rho$, $\rho^2$, $\rho^3$ or $\rho^4$ and using the freedom to exchange the resulting the (22) and (33) elements of $G_\ell$.

Given that we begin with relations of the form $G_\ell^5=(G_iG_\ell)^{m_i}=\mathbf{1}$, we seek embeddings in the polyhedral group $H(m,n)$, for which the abstract generators $s$ and $t$ satisfy the defining relation $s^2=t^{n}=(st)^{m}=1$.  Note that $H(m,n)=H(n,m)$ through a redefinition of the abstract generators $s$ and $t$.   The number of {\it finite} polyhedral groups is strongly constrained, as $H(m,n)$ is finite if and only if $2(n+m)>nm$ \cite{book:generator}.  These finite groups are $H(n,2)$, $H(3,3)$, $H(3,4)$, and $H(3,5)$, where $H(2,2)=K_4$, $H(n,2)=D_n\,(n\geq3)$, $H(3,3)=A_4$, $H(3,4)=S_4$, and $H(3,5)=A_5$.

Here we are considering $n=5$, and hence the finite group options are $D_n$ ($n\geq 3$) and $A_5$.  For the cases with $G_{\ell_1}$ and $G_{\ell_2}$, $m_i=3$ and $m_i=2$,  which corresponds to subgroups of $A_5$ and $D_5$, respectively.  For $m_i=3$ ({\it i.e.}, the group generated by $\{ G_1,\, G_{\ell_1}\}$ and the group generated by $\{G_2,\, G_{\ell_2}\}$), the generators obey $G_\ell^5=\mathbf{1}$, $G_i^2=\mathbf{1}$, and $(G_iG_\ell)^3=\mathbf{1}$.  Hence, according to Lagrange's theorem of group theory \cite{book:group}, the group order should be divisible by 30 (the least common multiple of 2, 3, and 5).  However, the maximal order of the nontrivial subgroups of $A_5$ is 12, and thus for $m_i=3$, the corresponding group is precisely $A_5$.  By similar reasoning, for $m_i=2$ the group is $D_5$.  Since $D_5$ is a subgroup of $A_5$, we conclude that the minimal family symmetry group for this class of solutions is indeed $A_5$. It is important to note that the groups generated for $m_i=5$ ({\it i.e.}, by $\{ G_2,\, G_{\ell_1}\}$ or by $\{G_1,\, G_{\ell_2}\}$) are the same ($A_5$) as the ones generated for the $m_i=3$ case given above.  For the remaining solutions, the fact that the $G_{\ell}$'s are related to $G_{\ell_1}$ or $G_{\ell_2}$ indicates that these cases generate identical groups.  The groups generated are either a 300-element group consisting of the elements of $A_5$,  $\rho A_5$,    $\rho^2 A_5$,  $\rho^3 A_5$, and  $\rho^4 A_5$,  or a 50-element group consisting of the elements of $D_5$,  $\rho D_5$,    $\rho^2 D_5$,  $\rho^3 D_5$, and  $\rho^4 D_5$.  Since $A_5$ is a subgroup of the larger 300-element group, the minimal family symmetry group is precisely $A_5$ for all cases.  As stated, it is minimal in the sense that any other viable horizontal symmetry group should contain $A_5$ as a subgroup to ensure that the full $K_4$ symmetry of the GR1 neutrino mass matrix is preserved.

We must also show that this is a faithful representation of the group.  To this end, we begin by expressing the elements of $A_5$ in terms of the generators $S$ and $T$, which satisfy the $A_5$ relations $S^2=T^5=(ST)^3=\mathbf{1}$; the results are shown for completeness in Section 3.  Taking the specific choice that $S=G_1$ and $T=\mathrm{diag}(1,\rho,\rho^4)$ (or equivalently $T=\mathrm{diag}(1,\rho^4,\rho)$),\footnote{Note that a similar choice of $S$ and $T$ was taken in  \cite{Feruglio:2011qq}.} it is straightforward to show that all of the representation matrices of the $A_5$ group elements are distinct. This ensures that we indeed have a faithful representation of the group.  With this choice of $S$ and $T$, $G_2$ and $G_3$ are given by
\begin{equation}
G_2=T^3ST^2ST^3,\qquad G_3=T^3ST^2ST^3S.
\label{strel}
\end{equation}
Note that we could also choose a representation in which $S=G_2$ and $T=\mathrm{diag}(1,\rho^2,\rho^3)$ (or $T=\mathrm{diag}(1,\rho^3,\rho^2)$), for which $G_1$ and $G_3$ are then given by Eq.~(\ref{strel}) with $G_2\rightarrow G_1$.

These results were obtained for the choice that $n=5$ ({\it i.e.}, $G_\ell^5=\mathbf{1}$).  One might ask the question as to whether higher values of $n$ can result in a smaller minimal family symmetry group than $A_5$.  We have explicitly checked that for a range of $n$ from 6 to 100, the resulting family symmetry group is finite if and only if $n=5k$, where $k$ is a positive integer.  Hence, $A_5$ is still the minimal group in these cases.  We did not consider the case of $n>100$ explicitly, but we can argue that the result should hold in the following way.   The order $j$ of $G_\ell$ should be a factor of $n$.  However, for $j\leq 100$, the horizontal group is finite if and only if $j$ is divisible by 5, while if $i>100$, the order of the group should also exceed 100. Hence, while it is not a mathematical proof, this indicates that the minimal family symmetry group is indeed $A_5$.  To obtain a more rigorous proof, one can follow the procedure of \cite{Lam:2008rs} of enumerating and rejecting other relevant finite groups.  This topic is left for future work.

For the case of GR2 mixing, we applied an identical procedure as outlined above based on the symmetry generators $G'_i$ given in Eq.~(\ref{11}), in which we begin by enumerating possible choices for $n$ and examining whether such cases can result in a finite family symmetry group with one or more of the $G'_i$'s.  In a systematic scan of the possible values of $n$ up to $n=100$, we were unable to find solutions for the integers $m_i$ such that $(G'_iG_\ell)^{m_i}=\mathbf{1}$, and hence the symmetry groups in these cases are infinite. It is worth noting, however, that just as in the case of GR1 mixing, one cannot rule out the possibility of a discrete group with a very large order.  Once again, we defer a more detailed study of this topic for future investigation. Note that it can be shown that the dihedral group $D_{10}$ used in \cite{GR2} only contains $G_\ell$ and $G'_3$ and not $G'_{1,2}$, and hence it is not minimal according to the criteria specified here.

For this reason, we now concentrate our attention on the case of GR1 mixing models based on the minimal family symmetry group of $A_5$ for the remainder of this work.   After  a brief summary of group theoretic aspects of $A_5$ given in the following section of our paper, we turn in Section 4 to the question of flavor model building based on $A_5$ that is predicated on our symmetry analysis.   The reader that wishes to turn directly to the issue of flavor model building should therefore skip Section 3 and proceed to Section 4.

\section{The discrete group $A_5$}
In this section, we enumerate basic group theoretic properties of $A_5$.  $A_5$ is the alternative group of order five; {\it i.e.}, it is the group of even permutations of five objects.   From the geometrical point of view, $A_5$ is the subgroup of the
three-dimensional rotation group that preserves the symmetries of the icosahedron.
The order of $A_5$ is 60; due to this relatively large value, the structure of $A_5$ is more complex than that of many other family symmetry groups such as $A_4$ and $S_4$ that are considered in the literature. The $A_5$ group elements can be generated by two generators $S$ and $T$ that obey the relations:
\begin{equation}
\label{eq:generator}S^2=T^5=(ST)^3=1.
\end{equation}
Without loss of generality, we can take \cite{book:generator}
\begin{equation}
S=(12)(45),~~~T=(13542),
\end{equation}
in which the cycle (13542) denotes the permutation
$(1,3,5,4,2)\rightarrow(3,5,4,2,1)$, and (12)(45) denotes
$(1,2,3,4,5)\rightarrow(2,1,3,5,4)$. The 60 elements of $A_5$ belong to five
conjugacy classes ${\cal C}_{i}$ ($i=1,\ldots 5$).  In terms of Schoenflies notation, in which $C_n^k$ labels rotations by $2\pi k/n$ and the number in front identifies the number of elements in each conjugacy class (and for $k=1$, the superscript is not written), ${\cal C}_1=1$, ${\cal C}_2= 15C_2$, ${\cal C}_3=20 C_3$, ${\cal C}_4=12 C_5$, and ${\cal C}_5=12 C_5^2$.  The group elements of each conjugacy class can be expressed in terms of $S$ and $T$ as follows:
\begin{eqnarray}
\nonumber&&{\cal C}_1:1,\\
\nonumber&&{\cal C}_2:ST^2ST^3S=(12)(34),~~TST^4=(13)(24),~~T^4(ST^2)^2=(14)(23),\\
\nonumber&&~~~~T^2ST^3=(12)(35),~~T^2ST^2ST^3S=(13)(25),~~ST^2ST=(15)(23),\\
\nonumber&&~~~~S=(12)(45),~~T^3 ST^2ST^3=(14)(25),~~T^3ST^2ST^3S=(15)(24),\\
\nonumber&&~~~~T^3ST^2=(13)(45),~~T^4ST^2ST^3S=(14)(35),~~TST^2S=(15)(34),\\
\nonumber&&~~~~ST^3ST^2S=(23)(45),~~T^4ST=(24)(35),~~(T^2S)^2T^4=(25)(34),\\
\nonumber&&{\cal C}_3:ST=(134),~~TS=(235),~~ST^4=(253),~~T^4S=(143),~~TST^3=(145),\\
\nonumber&&~~~~T^2ST^2=(243),~~T^2ST^4=(154),~~T^3ST=(125),~~T^3ST^3=(234),\\
\nonumber&&~~~~T^4ST^2=(152),~~TST^3S=(124),~~T^2ST^3S=(354),~~T^3ST^2S=(123), \\
\nonumber&&~~~~ST^2ST^3=(345),~~ST^3ST=(245),~~ST^3ST^2=(132),~~(T^2S)^2T^2=(135)\\
\nonumber&&~~~~T^2(T^2S)^2=(254),~~(ST^2)^2S=(153),~~(ST^2)^2T^2=(142),\\
\nonumber&&{\cal
C}_4:T=(13542),~~T^4=(12453),~~ST^2=(14235),~~T^2S=(13425),\\
\nonumber&&~~~~ST^3=(15243),~~T^3S=(15324),~~STS=(12345),~~TST=(15432),\\
\nonumber&&~~~~TST^2=(12534),~~T^2ST=(14523),~~T^3ST^4=(14352),~~T^4ST^3=(13254),\\
\nonumber&&{\cal
C}_5:T^2=(15234),~~T^3=(14325),~~ST^2S=(13524),~~ST^3S=(14253),\\
\nonumber&&~~~~(ST^2)^2=(12543),~~(T^2S)^2=(14532),~~(ST^3)^2=(12354),~~(T^3S)^2=(13452),\\
\nonumber&&~~~~(T^2S)^2T^3=(15423),~~T^3(ST^2)^2=(15342),~~T^3ST^2ST^4=(13245),\\
\nonumber&&~~~~T^4ST^2ST^3=(12435).
\end{eqnarray}
$A_5$ has 59 subgroups of order 1, 2, 3, 4, 5, 6, 10, 12 or 60, as follows:
\begin{enumerate}
\item{The trivial group that consists solely of the identity element.}
\item{The two-element subgroup generated by a double transposition:
}\vskip-0.3in
\begin{eqnarray}
\nonumber&& Z^{(1)}_2=\{1,ST^2ST^3S\}, Z^{(2)}_2=\{1,TST^4\},
Z^{(3)}_2=\{1,T^4(ST^2)^2\}, Z^{(4)}_2=\{1,T^2ST^3\},\\
\nonumber&&Z^{(5)}_2=\{1,T^2ST^2ST^3S\}, Z^{(6)}_2=\{1, ST^2ST\},
Z^{(7)}_2=\{1,S\}, Z^{(8)}_2=\{1,T^3ST^2ST^3\},\\
\nonumber&&Z^{(9)}=\{1,T^3ST^2ST^3S\}, Z^{(10)}_2=\{1,T^3ST^2\},
Z^{(11)}_2=\{1,T^4ST^2ST^3S\}, \\
\nonumber&&Z^{(12)}_2=\{1,TST^2S\}, Z^{(13)}_2=\{1, ST^3ST^2S\},
Z^{(14)}_2=\{1, T^4ST\}, Z^{(15)}_2=\{1,(T^2S)^2T^4\}.
\end{eqnarray}
\item{The three-element subgroup generated by a 3-cycle:}\vskip-0.3in
\begin{eqnarray}
\nonumber&&Z^{(1)}_3=\{1, T^3ST^2S, ST^3ST^2\}, Z^{(2)}_3=\{1,
TST^3S, (ST^2)^2T^2\}, Z^{(3)}_3=\{1, T^3ST, T^4ST^2\},\\
\nonumber&&Z^{(4)}_3=\{1, ST, T^4S\}, Z^{(5)}_3=\{1, (T^2S)^2T^2,
(ST^2)^2S\}, Z^{(6)}_3=\{1, TST^3, T^2ST^4\},\\
\nonumber&&Z^{(7)}_3=\{1, T^2ST^2, T^3ST^3\}, Z^{(8)}_3=\{1, TS,
ST^4\}, Z^{(9)}_3=\{1, ST^3ST, T^2(T^2S)^2\},\\
\nonumber&&Z^{(10)}_3=\{1, ST^2ST^3, T^2ST^3S\}.
\end{eqnarray}
\item{The four-element subgroup comprising of all double transitions on four of five objects, which is isomorphic to $K_4$:}\vskip-0.3in
\begin{eqnarray}
\nonumber&&K^{(1)}_4=\{1, ST^2ST^3S, TST^4, T^4(ST^2)^2\},
K^{(2)}_4=\{1, T^2ST^3, T^2ST^2ST^3S, ST^2ST\},\\
\nonumber&&K^{(3)}_4=\{1, S, T^3ST^2ST^3, T^3ST^2ST^3S\},
K^{(4)}_4=\{1, T^3ST^2, T^4ST^2ST^3S, TST^2S\},\\
\nonumber&&K^{(5)}_4=\{1,ST^3ST^2S, T^4ST, (T^2S)^2T^4\}.
\end{eqnarray}
\item{The five-element subgroup generated by a 5-cycle, which is isomorphic to
$Z_5$:}\vskip-0.3in
\begin{eqnarray}
\nonumber&&Z^{(1)}_5=\{1, STS, ST^2S, ST^3S, TST\},Z^{(2)}_5=\{1,
ST^3, T^2S, (ST^3)^2, (T^2S)^2\},\\
\nonumber&&Z^{(3)}_5=\{1, T^2ST, T^4ST^3, T^3(ST^2)^2,
T^4ST^2ST^3\}, Z^{(4)}_5=\{1, T, T^2, T^3, T^4\},\\
\nonumber&&Z^{(5)}_5=\{1, TST^2, T^3ST^4, (T^2S)^2T^3,
T^3ST^2ST^4\}, Z^{(6)}_5=\{1, ST^2, T^3S, (ST^2)^2, (T^3S)^2\}.
\end{eqnarray}
\item{The subgroup of order six in which three objects obey symmetric permutations, which is isomorphic to $S_3$:}
\vskip-0.3in
\begin{eqnarray}
\nonumber&&S^{(1)}_3=\{1, S, T^3ST^2, T^3ST^2S, ST^3ST^2,
ST^3ST^2S\}, \\
\nonumber&&S^{(2)}_3=\{1, T^2ST^3, T^4ST, TST^3S, (ST^2)^2T^2, T^4ST^2ST^3S\},\\
\nonumber&&S^{(3)}_3=\{1, T^3ST, T^4ST^2, TST^2S, (T^2S)^2T^4,
ST^2ST^3S\},\\
\nonumber&&S^{(4)}_3=\{1, ST, T^4S, (T^2S)^2T^4, T^3ST^2ST^3,
T^2ST^2ST^3S\},\\
\nonumber&&S^{(5)}_3=\{1, TST^4, T^4ST, (T^2S)^2T^2, (ST^2)^2S,
T^3ST^2ST^3S\},\\
\nonumber&&S^{(6)}_3=\{1,TST^3, T^2ST^4, ST^2ST, T^4(ST^2)^2,
ST^3ST^2S\},\\
\nonumber&&S^{(7)}_3=\{1, T^2ST^2, T^3ST^3, ST^2ST, TST^2S,
T^3ST^2ST^3S\},\\
\nonumber&&S^{(8)}_3=\{1, TS, ST^4, T^4(ST^2)^2, T^3ST^2ST^3,
T^4ST^2ST^3S\},\\
\nonumber&&S^{(9)}_3=\{1, TST^4, T^3ST^2, ST^3ST, T^2(T^2S)^2,
T^2ST^2ST^3S\},\\
\nonumber&&S^{(10)}_3=\{1, S, T^2ST^3, ST^2ST^3, T^2ST^3S,
ST^2ST^3S\}.
\end{eqnarray}
\item{The order ten subgroup generated by a 5-cycle and a double transition which conjugates it to its inverse, which is isomorphic to
$D_{5}$:}\vskip-0.3in
\begin{eqnarray}\
\nonumber&&D^{(1)}_{5}=\{1, STS, ST^2S, ST^3S, TST, T^2ST^3,
T^3ST^2, T^4(ST^2)^2, (T^2S)^2T^4, T^3ST^2ST^3S\},\\
\nonumber&&D^{(2)}_{5}=\{1, ST^3, T^2S, (ST^3)^2, (T^2S)^2,
T^3ST^2, T^4ST, ST^2ST, ST^2ST^3S, T^3ST^2ST^3 \},\\
\nonumber&&D^{(3)}_{5}=\{1, S, TST^4, T^2ST, T^4ST^3,ST^2ST,
T^3(ST^2)^2, (T^2S)^2T^4, T^4ST^2ST^3, T^4ST^2ST^3S\},\\
\nonumber&&D^{(4)}_{5}=\{1, T, T^2, T^3, T^4, ST^2ST^3S, ST^3ST^2S,
T^2ST^2ST^3S, T^3ST^2ST^3S, T^4ST^2ST^3S\},\\
\nonumber&&D^{(5)}_{5}=\{1, S, TST^2, T^3ST^4, T^4ST, TST^2S,
(T^2S)^2T^3, T^4(ST^2)^2, T^3ST^2ST^4, T^2ST^2ST^3S\},\\
\nonumber&&D^{(6)}_{5}=\{1, ST^2, T^3S, (ST^2)^2, (T^3S)^2, TST^4,
T^2ST^3, TST^2S, ST^3ST^2S, T^3ST^2ST^3\}.
\end{eqnarray}
\item{The twelve-element subgroup of permutations of any four of the five objects, which is isomorphic to $A_4$:}\vskip-0.3in
\begin{eqnarray}
\nonumber&&A^{(1)}_4=\{1, ST, T^4S, TST^4, T^2ST^2, T^3ST^3,
ST^3ST^2, TST^3S, T^3ST^2S, (ST^2)^2T^2, T^4(ST^2)^2,\\
\nonumber&&\hskip0.4inST^2ST^3S\},\\
\nonumber&&A^{(2)}_4=\{1,ST^4, TS, T^2ST^3, T^3ST, T^4ST^2, ST^2ST,
ST^3ST^2, T^3ST^2S, (T^2S)^2T^2, (ST^2)^2S, \\
\nonumber&&\hskip0.4inT^2ST^2ST^3S \},\\
\nonumber&&A^{(3)}_4=\{1, S, TST^3, T^2ST^4, T^3ST, T^4ST^2, ST^3ST,
TST^3S, (ST^2)^2T^2, T^2(T^2S)^2, T^3ST^2ST^3,\\
\nonumber&&\hskip0.4inT^3ST^2ST^3S\},\\
\nonumber&&A^{(4)}_4=\{1, ST, T^4S, TST^3, T^2ST^4, T^3ST^2,
ST^2ST^3, TST^2S, T^2ST^3S, (ST^2)^2S, (T^2S)^2T^2,\\
\nonumber&&\hskip0.4inT^4ST^2ST^3S\},\\
\nonumber&&A^{(5)}_4=\{1, ST^4, TS, T^2ST^2, T^3ST^3, T^4ST,
ST^2ST^3, ST^3ST, T^2ST^3S, T^2(T^2S)^2, (T^2S)^2T^4,\\
\nonumber&&\hskip0.4inST^3ST^2S\}.
\end{eqnarray}
\item{The full group.}
\end{enumerate}
\begin{table}[t]
\begin{center}
\begin{tabular}{|c|c|c|c|c|c|}\hline\hline
   &\multicolumn{5}{c|}{Conjugacy Classes}\\\cline{2-6}
   &${\cal C}_1$&${\cal C}_2$&${\cal C}_3$&${\cal C}_4$&${\cal
C}_5$ \\\hline

$\mathbf{1}$&1&1&1&1&1\\\hline

$\mathbf{3}$&3&-1&0& $\phi_g$ & $1-\phi_g$
\\\hline

$\mathbf{3}'$&3&-1&0&$1-\phi_g$& $\phi_g$ \\\hline

$\mathbf{4}$&4&0&1&-1 &-1 \\\hline

$\mathbf{5}$&5&1&-1&0  & 0 \\\hline\hline

\end{tabular}
\caption{\label{tab:character} The character table of the $A_5$ group,
in which the conjugacy classes written in Schoenflies notation are ${\cal C}_1=1$, ${\cal C}_2= 15C_2$, ${\cal C}_3=20 C_3$, ${\cal C}_4=12 C_5$, and ${\cal C}_5=12 C_5^2$.}
\end{center}
\end{table}
For completeness, we now summarize a number of basic group theoretic facts about $A_5$ that are needed for flavor model building,  which have previously been presented in the literature.  We begin by recalling that $A_5$ has five irreducible representations,  $\mathbf{1}$, $\mathbf{3}$, $\mathbf{3}'$, $\mathbf{4}$ and $\mathbf{5}$.  The $A_5$ character table is presented in Table \ref{tab:character}, and the products of irreducible representations, which are easily determined from the character table, are listed below:
\begin{eqnarray}
\nonumber&&\mathbf{1}\otimes R=R\otimes\mathbf{1}=R,~~~\mathbf{3}\otimes\mathbf{3}=\mathbf{1}\oplus\mathbf{3}\oplus\mathbf{5},~~~\mathbf{3}'\otimes\mathbf{3}'=\mathbf{1}\oplus\mathbf{3}'\oplus\mathbf{5}, ~~~\mathbf{3}\times\mathbf{3}'=\mathbf{4}\oplus\mathbf{5},\\
\nonumber&&\mathbf{3}\otimes\mathbf{4}=\mathbf{3}'\oplus\mathbf{4}\oplus\mathbf{5},~~~\mathbf{3}'\otimes\mathbf{4}=\mathbf{3}\oplus\mathbf{4}\oplus\mathbf{5},~~~\mathbf{3}\otimes\mathbf{5}=\mathbf{3}\oplus\mathbf{3}'\oplus\mathbf{4}\oplus\mathbf{5},\\
\nonumber&&\mathbf{3}'\otimes\mathbf{5}=\mathbf{3}\oplus\mathbf{3}'\oplus\mathbf{4}\oplus\mathbf{5},~~~\mathbf{4}\otimes\mathbf{4}=\mathbf{1}\oplus\mathbf{3}\oplus\mathbf{3}'\oplus\mathbf{4}\oplus\mathbf{5},~~~\mathbf{4}\otimes\mathbf{5}=\mathbf{3}\oplus\mathbf{3}'\oplus\mathbf{4}\oplus\mathbf{5}_1\oplus\mathbf{5}_2,\\ &&\mathbf{5}\otimes\mathbf{5}=\mathbf{1}\oplus\mathbf{3}\oplus\mathbf{3}'\oplus\mathbf{4}\oplus\mathbf{4}\oplus\mathbf{5}_1\oplus\mathbf{5}_2.
\end{eqnarray}
Here $R$ indicates any $A_{5}$ irreducible representation and $\mathbf{5}_{1,2}$ stand for the two $\mathbf{5}$ representations that appear in the Kronecker products. In the Appendix, we list the explicit representation matrices of the generators $S$ and $T$ for each of the irreducible representations, and provide the corresponding Clebsch-Gordan coefficients.

\section{\label{sec:model1} $A_5$ Flavor Models with Triplet Embeddings}

Guided by the above symmetry analysis, we now turn to the issue of lepton flavor model-building based on $A_5$.  Lepton flavor models have previously been constructed within this framework \cite{A5everettstuart,Feruglio:2011qq}.  In this work, we specifically seek natural models of GR1 mixing in which the SM leptons are embedded in the {\it triplet} representations (as opposed to singlet representations, as done in \cite{Feruglio:2011qq}), and in which issues of vacuum alignment can be achieved naturally.  In this section, we will present two $A_5$ flavor models that result in GR1 mixing: one in which the family symmetry is broken to $Z_5$ and $K_4$ in the charged lepton and neutrino sectors, respectively, and one in which the $Z_5$ symmetry is itself broken in the charged lepton sector. The details of each scenario will be described below.  However, we begin by reviewing basic aspects of flavor model building within the framework of $A_5$.

The basic paradigm is that the full family symmetry under which the matter fields transform is realized at some high energy scale $\Lambda$, and  is then spontaneously broken to subgroups by flavon fields, which are scalar multiplets with nonvanishing vacuum expectation values (VEVs).  For the leptons, the subgroup that is preserved in the charged lepton mass matrix should be different than the surviving subgroup in the neutrino sector.  This misalignment between the symmetry breaking patterns results in mass-independent textures (such as GR1 or tri-bimaximal mixing) at leading order (LO).  Hence, group theoretical considerations of the family symmetry breaking chains are critical for guiding model-building efforts.

As we have demonstrated in section \ref{sec:minimal}, the GR1 mixing pattern is naturally produced in $A_5$ models when the $A_5$ symmetry is broken to the $Z_5$ subgroup generated by $T$ in the charged lepton sector, and broken down to the $K_4$ subgroup comprised of $S$, $T^3ST^2ST^3$, $T^3ST^2ST^3S$ and the identity in the neutrino sector.   Since the lepton masses are generated after the spontaneous breaking of the flavor symmetry, the flavon VEVs must be invariant under the corresponding subgroups, which strongly constrains the vacuum alignment in the flavon field sector.  More precisely, for the flavon field $\varphi_T$ associated with the charged leptons at LO, we see from the explicit representation matrices shown in the Appendix that the invariance under the action of $T$ means we should restrict ourselves to the following set of flavon field VEVs:
\begin{eqnarray}
\langle\varphi_{T}\rangle\propto\left\{\begin{array}{ll}(1,0,0),&~~~~~~\varphi_{T}\sim\mathbf{3}\\
(1,0,0),&~~~~~~\varphi_{T}\sim\mathbf{3}'\\
(0,0,0,0),&~~~~~~\varphi_{T}\sim\mathbf{4}\\
(1,0,0,0,0),&~~~~~~\varphi_{T}\sim\mathbf{5}.\\
\end{array}\right.
\end{eqnarray}
Similar reasoning for the flavon field $\varphi_S$ associated with the $K_4$ subgroup of the neutrino sector, its vacuum configuration is determined to be
\begin{equation}
\langle\varphi_S\rangle=\left\{\begin{array}{ll}(1,1,1,1)v,&~~~~~~\varphi_S\sim\mathbf{4}\\
(\sqrt{\frac{2}{3}}(v_2+v_3),v_2,v_3,v_3,v_2),&~~~~~~\varphi_S\sim\mathbf{5},
\end{array}\right.
\end{equation}
in which $v$, $v_2$ and $v_3$ are arbitrary parameters. It is straightforward to see that if $\varphi_S$ is a $A_5$ triplet $\mathbf{3}$ or $\mathbf{3}'$, there is no vacuum alignment that maintains the $K_4$ symmetry, although such representations can break $A_5$ to the  $Z_2$ subgroup generated by $S$ and $T^3ST^2ST^3$ or $T^3ST^2ST^3S$ separately.

\begin{table}[t]
\begin{center}
\begin{tabular}{|c|c|c|c|c||c|c|c|c|c||c|c|c|c|c|c|c|}\hline\hline
Fields & $L$ & $E^c$ & $N^{c}$ &$h_{u,d}$ & $\xi$  &  $\chi$  & $\phi$ &  $\Delta$ & $\varphi$ & $\chi^0$ & $\phi^0$ &  $\Delta^0$  &$\psi^{0}$
\\\hline

$A_5$  & $\mathbf{3}$ &  $\mathbf{3}$  &  $\mathbf{3}'$ & $\mathbf{1}$  & $\mathbf{1}$  &  $\mathbf{3}$  &  $\mathbf{5}$ & $\mathbf{4}$  &  $\mathbf{5}$  &  $\mathbf{3}$  & $\mathbf{5}$ &  $\mathbf{4}$ & $\mathbf{3}'$  \\

$Z_5$ &  $\rho^3$ &  1 &  $\rho^2$  &  1  &  $\rho^2$  &  $\rho^2$  &  $\rho^2$  &   1  &  $\rho$  & $\rho$ &  $\rho$ &  1  &  $\rho^4$  \\  \hline\hline

\end{tabular}
\caption{\label{tab:field1} The transformation properties of the matter fields, the electroweak Higgs doublets, the flavon fields and the driving fields under the flavor symmetry $A_{5}\times Z_5$, where $\rho=e^{2\pi i/5}$, $L=(\ell_1,\ell_2,\ell_3)$ are the lepton doublet, $E^c=(e^{c}, \tau^{c}, \mu^{c})$ is the charged leptons, and $N^{c}=(\nu^{c}_1,\nu^{c}_2,\nu^{c}_3)$ are the right-handed neutrinos.}
\end{center}
\end{table}
\subsection{Model 1}
Armed with this knowledge, we now present the first model (Model 1), in which the full flavor symmetry is $A_5\times Z_5$. The particle content is shown in Table \ref{tab:field1}.  In keeping with our goal of triplet embeddings, the lepton doublets  $L=(\ell_1,\ell_2,\ell_3)$ and right-handed (RH) charged leptons $E^{c}=(e^{c},\tau^{c},\mu^{c})$ are both assigned to $A_5$ triplet ($\mathbf{3}$) representations. We note that $\tau^{c}$ and $\mu^{c}$ are taken to be the second and third components respectively so that the LO charged lepton mass matrix is diagonal, as will be seen below. The neutrino masses are generated via the Type I seesaw mechanism, in which the right-handed neutrinos $N^{c}=(\nu^{c}_1,\nu^{c}_2,\nu^{c}_3)$ transform as a $\mathbf{3}'$.  The flavon sector consists of the following set of fields:  a singlet $\xi$, triplet  $\chi$, and a 5-plet $\phi$, which couple to the charged leptons, and the 4-plet $\Delta$ and additional 5-plet $\varphi$, which couple to the neutrino sector. We use the now-standard supersymmetric driving field method to arrange the vacuum alignment, in which a continuous $U(1)_R$ symmetry related to $R-$parity is introduced under which the matter fields have a  $+1$ $R-$charge while the electroweak Higgs and flavon fields are uncharged.  Upon the introduction of so-called driving fields that carry a $U(1)_R$ charge of $+2$, we see that the flavon sector superpotential self-interactions must consist of terms in which the driving fields appear linearly. Hence, the minimization of the flavon potential can be achieved simply by ensuring that the F-terms of the driving fields vanish at the minimum.  In our model, we require four driving fields:  two triplets, one $\mathbf{3}$ ($\chi^0$) and one $\mathbf{3'}$ ($\psi^0$), one $\mathbf{4}$ ($\Delta^0$), and one $\mathbf{5}$ ($\phi^0$).  With this particle content, the superpotential $w$ takes the following form at leading order:
\begin{equation}
w=w_{\ell}+w_{\nu}+w_d,
\end{equation}
in which
\begin{eqnarray}
&&w_{\ell}=\frac{y_{\ell_1}}{\Lambda}(E^{c}L)_{\mathbf{1}}\xi h_d+\frac{y_{\ell_2}}{\Lambda}((E^{c}L)_{\mathbf{3}}\chi)_{\mathbf{1}}h_d+\frac{y_{\ell_3}}{\Lambda}((E^cL)_{\mathbf{5}}\phi)_{\mathbf{1}}h_d\\
\label{eq:nu}&&w_{\nu}=\frac{y_{\nu}}{\Lambda}((N^{c}L)_{\mathbf{4}}\Delta)_{\mathbf{1}}h_u+\frac{1}{2}x((N^{c}N^{c})_{\mathbf{5}}\varphi)_{\mathbf{1}}\\
&&w_d=f_1(\chi^0\chi)_{\mathbf{1}}\xi+f_2(\chi^0(\chi\phi)_{\mathbf{3}})_{\mathbf{1}}+{g_1(\phi^0\phi)_{\mathbf{1}}\xi}+g_2(\phi^0(\chi\chi)_{\mathbf{5}})_{\mathbf{1}}+g_3(\phi^0(\chi\phi)_{\mathbf{5}})_{\mathbf{1}}\\
\nonumber&&~~~~~+g_4(\phi^0(\phi\phi)_{\mathbf{5}_1})_{\mathbf{1}}+g_5(\phi^0(\phi\phi)_{\mathbf{5}_2})_{\mathbf{1}}+M_{\Delta}(\Delta^0\Delta)_{\mathbf{1}}+h_1(\Delta^0(\Delta\Delta)_{\mathbf{4}})_{\mathbf{1}}+h_2(\psi^0(\Delta\varphi)_{\mathbf{3}'})_{\mathbf{1}},
\end{eqnarray}
in which the subscripts $\mathbf{1}$, $\mathbf{3}$, $\mathbf{3}'$, $\mathbf{4}$ and $\mathbf{5}$ denote $A_5$ contractions.  In the above, we have neglected to write higher dimensional operators, which are suppressed by additional powers of the high (cutoff) scale $\Lambda$. We note that the combinations $(\chi^{0}(\chi\chi)_{\mathbf{3}})_{\mathbf{1}}$ and $(\chi^0(\phi\phi)_{\mathbf{3}})_{\mathbf{1}}$ are omitted here since the contractions $(\chi\chi)_{\mathbf{3}}$ and $(\phi\phi)_{\mathbf{3}}$ vanish.

The F-terms of the driving fields take the form
\begin{eqnarray}
\nonumber&&\frac{\partial w_d}{\partial\chi^0_1}=f_1\chi_1\xi+f_2(-2\chi_1\phi_1+\sqrt{3}\,\chi_2\phi_5+\sqrt{3}\,\chi_3\phi_2)=0\\
\nonumber&&\frac{\partial w_d}{\partial\chi^0_2}=f_1\chi_3\xi+f_2(\sqrt{3}\,\chi_1\phi_5-\sqrt{6}\,\chi_2\phi_4+\chi_3\phi_1)=0\\
\nonumber&&\frac{\partial w_d}{\partial\chi^0_3}=f_1\chi_2\xi+f_2(\sqrt{3}\,\chi_1\phi_2+\chi_2\phi_1-\sqrt{6}\,\chi_3\phi_3)=0\\
\nonumber&&\frac{\partial w_d}{\partial\phi^0_1}=g_1\phi_1\xi+2g_2(\chi^2_1-\chi_2\chi_3)+\sqrt{3}\,g_3(\chi_2\phi_5-\chi_3\phi_2)+2g_4(\phi^2_1+\phi_2\phi_5-2\phi_4\phi_4)\\
\nonumber&&\quad\quad+2g_5(\phi^2_1-2\phi_2\phi_5+\phi_3\phi_4)=0\\
\nonumber&&\frac{\partial w_d}{\partial\phi^0_2}=g_1\phi_5\xi-2\sqrt{3}\,g_2\chi_1\chi_3+g_3(\chi_1\phi_5+\sqrt{2}\,\chi_2\phi_4+\sqrt{3}\,\chi_3\phi_1)+2g_4(\phi_1\phi_5+\sqrt{6}\,\phi_2\phi_4)\\
\nonumber&&\quad\quad+g_5(-4\phi_1\phi_5+\sqrt{6}\,\phi^2_3)=0
\end{eqnarray}
\begin{eqnarray}
\nonumber&&\frac{\partial w_d}{\partial\phi^0_3}=g_1\phi_4\xi+\sqrt{6}\,g_2\chi^2_3+g_3(2\chi_1\phi_4+\sqrt{2}\,\chi_3\phi_5)+g_4(-4\phi_1\phi_4+\sqrt{6}\,\phi^2_5)\\
\nonumber&&\quad\quad+2g_5(\phi_1\phi_4+\sqrt{6}\,\phi_2\phi_3)=0\\
\nonumber&&\frac{\partial w_d}{\partial\phi^0_4}=g_1\phi_3\xi+\sqrt{6}\,g_2\chi^2_2+g_3(-2\chi_1\phi_3-\sqrt{2}\,\chi_2\phi_2)+g_4(-4\phi_1\phi_3+\sqrt{6}\,\phi^2_2)\\
\nonumber&&\quad\quad+2g_5(\phi_1\phi_3+\sqrt{6}\,\phi_4\phi_5)=0\\
\nonumber&&\frac{\partial w_d}{\partial \phi^0_5}=g_1\phi_2\xi-2\sqrt{3}\,g_2\chi_1\chi_2+g_3(-\chi_1\phi_2-\sqrt{3}\,\chi_2\phi_1-\sqrt{2}\,\chi_3\phi_3)+2g_4(\phi_1\phi_2+\sqrt{6}\,\phi_3\phi_5)\\
&&\quad\quad+g_5(-4\phi_1\phi_2+\sqrt{6}\,\phi^2_4)=0.
\label{ftermsmodel1}
\end{eqnarray}
A solution to these relations is:
\begin{equation}
\label{eq:vacuum1}\langle\xi\rangle=v_{\xi},~~~\langle\chi\rangle=(v_{\chi},0,0),~~~\langle\phi\rangle=(v_{\phi},0,0,0,0),
\end{equation}
in which the VEVs obey the relations
\begin{equation}
v_{\phi}=\frac{f_1}{2f_2}v_{\xi},~~~g_1v_{\phi}v_{\xi}+2g_2v^2_{\chi}+2(g_4+g_5)v^2_{\phi}=0, 
\end{equation}
with $v_{\xi}$ undetermined. The vacuum expectation values of $\Delta$ and $\varphi$, which give rise to GR1 mixing in the neutrino sector, are determined by the F-terms of the associated driving fields as follows:
\begin{eqnarray}
\nonumber&&\frac{\partial w_d}{\partial\Delta^0_1}=M_{\Delta}\Delta_4+h_1(2\Delta_1\Delta_3+\Delta^2_2)=0\\
\nonumber&&\frac{\partial w_d}{\partial\Delta^0_2}=M_{\Delta}\Delta_{3}+h_1(2\Delta_1\Delta_2+\Delta^2_4)=0\\
\nonumber&&\frac{\partial w_d}{\partial\Delta^0_{3}}=M_{\Delta}\Delta_{2}+h_1(\Delta^2_1+2\Delta_3\Delta_4)=0\\
\nonumber&&\frac{\partial w_d}{\partial\Delta^{0}_4}=M_{\Delta}\Delta_1+h_1(2\Delta_2\Delta_4+\Delta^2_3)=0
\end{eqnarray}
\begin{eqnarray}
\nonumber&&\frac{\partial w_d}{\partial\psi^0_1}=\sqrt{2}\,h_2(\Delta_1\varphi_5+2\Delta_2\varphi_4-2\Delta_3\varphi_3-\Delta_4\varphi_2)=0\\
\nonumber&&\frac{\partial w_d}{\partial\psi^0_2}=h_2(-2\Delta_1\varphi_3+\Delta_2\varphi_2+\sqrt{6}\,\Delta_3\varphi_1-3\Delta_4\varphi_5)=0\\
&&\frac{\partial w_d}{\partial\psi^0_3}=h_2(3\Delta_1\varphi_2-\sqrt{6}\,\Delta_2\varphi_1-\Delta_3\varphi_5+2\Delta_4\varphi_4)=0,
\end{eqnarray}
which admit the solution
\begin{equation}
\label{eq:vacuum2}\langle\Delta\rangle=(1,1,1,1)v_{\Delta},~~~\langle\varphi\rangle=(\sqrt{\frac{2}{3}}(v_2+v_3),v_2,v_3,v_3,v_2)
\end{equation}
where $v_{\Delta}=-M_{\Delta}/(3h_1)$ and $v_{2,3}$  are undetermined. The auxiliary symmetry $Z_5$ ensures that the flavons $\xi$, $\chi$ and $\phi$ which appear in $w_{\ell}$ don't contribute to $w_{\nu}$ and vice versa at LO.  With this set of flavon VEVs, $A_5$ is broken to $Z_5$ and $K_4$ in the charged lepton and neutrino sectors, respectively. After flavor and electroweak symmetry breaking, $w_{\ell}$ results in the following diagonal charged lepton mass matrix:
\begin{equation}
m_{\ell}=\left(\begin{array}{ccc}
y_{\ell_1}\frac{v_{\xi}}{\Lambda}+2y_{\ell_3}\frac{v_{\phi}}{\Lambda} &  0  & 0\\
0&y_{\ell_1}\frac{v_{\xi}}{\Lambda}-y_{\ell_2}\frac{v_{\chi}}{\Lambda}-y_{\ell_3}\frac{v_{\phi}}{\Lambda}&0\\
0&0& y_{\ell_1}\frac{v_{\xi}}{\Lambda}+y_{\ell_2}\frac{v_{\chi}}{\Lambda}-y_{\ell_3}\frac{v_{\phi}}{\Lambda}
\end{array}\right)v_d,
\end{equation}
in which $v_d=\langle h_d\rangle$ is the VEV of the electroweak Higgs field $h_d$. The three independent combinations $y_{\ell_1}v_{\xi}v_d/\Lambda$, $y_{\ell_2}v_{\chi}v_d/\Lambda$ and $y_{\ell_3}v_{\phi}v_d/\Lambda$ can be expressed in terms of the charged lepton masses as follows:
\begin{eqnarray}
\nonumber&&y_{\ell_1}\frac{v_{\xi}}{\Lambda}v_d=\frac{1}{3}(m_e+m_{\mu}+m_{\tau})\\
\nonumber&&y_{\ell_2}\frac{v_{\chi}}{\Lambda}v_d=\frac{1}{2}(m_{\tau}-m_{\mu})\\
&&y_{\ell_3}\frac{v_{\phi}}{\Lambda}v_d=\frac{1}{6}(2m_e-m_{\mu}-m_{\tau}).
\end{eqnarray}
We see that $y_{\ell_1}v_{\xi} : y_{\ell_2}v_{\chi}: y_{\ell_3}v_{\phi}\approx 2:3:-1$. Hence, to reproduce the observed mass hierarchies of the charged leptons, additional fine tuning of the parameters $y_{\ell_i}$ is required.

Turning now to the neutrino sector, the Dirac neutrino mass matrix as obtained from the first term of Eq.~(\ref{eq:nu}) is given by
\begin{equation}
m_D=y_{\nu}\left(\begin{array}{ccc}
0 & \sqrt{2}  & \sqrt{2} \\
-\sqrt{2}  & -1&  1 \\
-\sqrt{2}  & 1 & -1
\end{array}\right)\frac{v_{\Delta}}{\Lambda}v_u,
\end{equation}
in which $v_u=\langle h_u\rangle$. The second term in Eq.~(\ref{eq:nu}) results in the Majorana mass matrix
\begin{equation}
m_M=x\left(\begin{array}{ccc}
2\sqrt{\frac{2}{3}}\,(v_2+v_3) & -\sqrt{3}\,v_3  & -\sqrt{3}\,v_3\\
-\sqrt{3}\,v_3 & \sqrt{6}\,v_2  & -\sqrt{\frac{2}{3}}\,(v_2+v_3)\\
-\sqrt{3}\,v_3  & -\sqrt{\frac{2}{3}}\,(v_2+v_3) & \sqrt{6}\,v_2
\end{array}\right).
\end{equation}
The light neutrino mass matrix $m_{\nu}=-m^{T}_Dm^{-1}_Mm_D$ is exactly diagonalized by the GR1 mixing matrix $U_{{\rm GR}1}$ via
$U^{T}_{{\rm GR}1}m_{\nu}U_{{\rm GR}1}=\mathrm{diag}(m_1,m_2,m_3)$, in which $U_{{\rm GR}1}$ is given by Eq.~(\ref{mnspgen}) with $\cot\theta_{12}=\phi_g$.  The mass eigenvalues are
\begin{equation}
\label{eq:mass1}m_1=-\frac{4\sqrt{6}\,y^2_{\nu}v^2_{\Delta}}{(a-b)\Lambda}\frac{v^2_u}{\Lambda},~~~m_2=-\frac{4\sqrt{6}\,y^2_{\nu}v^2_{\Delta}}{(a+b)\Lambda}\frac{v^2_u}{\Lambda},~~~m_3=-\frac{2\sqrt{6}\,y^2_{\nu}v^2_{\Delta}}{a\Lambda}\frac{v^2_u}{\Lambda},
\end{equation}
in which the parameters $a$ and $b$ are defined as
\begin{equation}
a=x(4v_2+v_3),~~~~b=3\sqrt{5}\,xv_3.
\end{equation}
The neutrino mass spectrum is strongly constrained in the present model, as it is determined by three independent real parameters: the absolute values of $a$ and $b$ and the relative phase between them. In particular, note that the following sum rule is obeyed:
\begin{equation}
\frac{1}{m_1}+\frac{1}{m_2}=\frac{1}{m_3}.
\end{equation}
This sum rule has also been obtained in a number of other models \cite{Barry:2010yk}.
The mass spectrum can either be a  normal hierarchy (NH) or inverted hierarchy (IH). Taking into account the measured solar and atmospheric mass-squared differences, we obtain the following limits for the lightest neutrino mass:
\begin{eqnarray}
\nonumber&|m_1|\geq0.011\,\mathrm{eV},  & \quad\quad\mathrm{NH}\\
&|m_3|\geq0.028\,\mathrm{eV},  & \quad\quad\mathrm{IH}.
\end{eqnarray}
We note that an analytical formula for the lower limits of neutrino masses that obey this sum rule has been determined in \cite{Barry:2010yk}.
In addition, this model has definite predictions for the Majorana phases $\alpha_{13}$ and $\alpha_{23}$ (using the standard parameterization \cite{pdg} of the MNSP mixing matrix). These phases are determined by the phase differences between the complex masses of Eq.~(\ref{eq:mass1}). After some algebraic manipulation, the following relations can be obtained:
\begin{eqnarray}
\nonumber&&\cos\alpha_{13}=\frac{1}{2|m_1||m_2|^2|m_3|}\big[|m_1|^2|m_2|^2-(|m_1|^2-|m_2|^2)|m_3|^2\big]\\
&&\cos\alpha_{23}=\frac{1}{2|m_1|^2|m_2||m_3|}\big[|m_1|^2|m_2|^2+(|m_1|^2-|m_2|^2)|m_3|^2\big].
\end{eqnarray}
Finally, we comment that in this model,  the neutrinoless double beta decay parameter $m_{ee}$  can also be expressed in terms of the light neutrino masses and the golden ratio $\phi_g$ as follows:
\begin{equation}
|m_{ee}|=\frac{1}{\sqrt{5}\,|m_3|}\big[|m_1|^2|m_2|^2+\phi_g |m_1|^2|m_3|^2-\frac{1}{\phi_g}|m_2|^2|m_3|^2\big]^{1/2}.
\end{equation}
The lower bound of $|m_{ee}|$ is found to be about 4 meV and 52 meV for the NH and IH cases, respectively. While the GR1 mixing pattern emerges naturally within this model, we must tune the Yukawa couplings $y_{\ell_i}$ to fit the charged lepton masses. Further fine tuning is required if subleading corrections are included. We note that this kind of fine tuning does not contradict the logic of Section 2, which only focuses on flavor mixing. In what follows, we aim to improve on this scenario by proposing another $A_5$ model that overcomes this defect by breaking the residual $Z_5$ symmetry in the charged lepton sector.\\

\subsection{Model 2}
\subsubsection{LO predictions}
To improve on the previous model, note that there are four possible assignments for the lepton doublets and the charged leptons: (i) both can be $\mathbf{3}$ or $\mathbf{3}'$ representations, (ii) one is a $\mathbf{3}$ and the other is a $\mathbf{3}'$, (iii)  one is a $\mathbf{3}$ or $\mathbf{3}'$ while the other is an $A_5$ singlet, and (iv) both are $A_5$ singlets.  We set aside case (iv), in which $A_5$ does not play a fundamental role in the lepton sector.  In case (ii), which was considered in \cite{A5everettstuart}, it is in general difficult to avoid nonvanishing off-diagonal elements in the charged lepton mass matrix, which will result in deviations from GR1 mixing.  Case (iii) was considered by \cite{Feruglio:2011qq}, in which an additional $Z_n$ was introduced to ensure a diagonal charged lepton mass matrix; in this model, the necessary vacuum alignment is not realized consistently within the driving fields method.

\begin{table}[hptb]
\begin{center}
\begin{tabular}{|c|c|c|c|c||c|c|c|c||c|c|c|c|}\hline\hline
Fields & $L$ & $E^c$ & $N^{c}$ &$h_{u,d}$ &   $\chi$  & $\phi$ &  $\zeta$ & $\varphi$ & $\xi^0$ &  $\rho^0$  & $\chi^0$ &  $\varphi^{0}$
\\\hline

$A_5$  & $\mathbf{3}$ &  $\mathbf{3}$  &  $\mathbf{3}$ & $\mathbf{1}$  &  $\mathbf{5}$  &  $\mathbf{5}$ & $\mathbf{1}$  &  $\mathbf{5}$  &  $\mathbf{1}$  & $\mathbf{3}'$  & $\mathbf{5}$  &   $\mathbf{5}$  \\

$Z_3$ & 1 & $\omega^2$  & 1 & 1 & $\omega^2$  &  $\omega$ & 1 & 1 & 1 & 1 & $\omega$   &  1  \\

$Z_3$ & $\omega^2$  & $\omega$  &  1  & 1 & 1 & 1 &   $\omega$  &  $\omega$ & 1  & 1 & 1 &  $\omega$  \\\hline\hline

\end{tabular}
\caption{\label{tab:field2}
The transformation properties of the matter fields, the Higgs doublets, the flavon fields and the driving fields under the flavor symmetry  $A_{5}\times Z_3 \times Z_3$, in which $\omega=e^{2\pi i/3}$, $L=(\ell_1,\ell_2,\ell_3)$ are the lepton doublet, $E^c=(e^{c}, \mu^{c}, \tau^{c})$ are the charged leptons, and $N^{c}=(\nu^{c}_1,\nu^{c}_2,\nu^{c}_3)$ are the right-handed neutrinos.}
\end{center}
\end{table}

Given that we wish to focus on triplet embeddings that result in GR1 mixing without excessive fine-tuning, we thus focus on case (i), and assign the SM leptons to $\mathbf{3}$ representations ($\mathbf{3'}$ representations  give essentially the same results).  The family symmetry of our second model is $A_5\times Z_3\times Z_3$, and the particle content is given in Table \ref{tab:field2}.  We see that in this case the flavon fields are only in $\mathbf{1}$ ($\zeta$) and $\mathbf{5}$ ($\chi$, $\phi$, $\varphi$) representations.  Four driving fields are needed: one singlet ($\xi^0$), one $\mathbf{3}'$ ($\rho^0$), and two $\mathbf{5}$'s ($\chi^0$ and $\varphi^0$).  As before, the superpotential of the model takes the form
\begin{eqnarray}
w=w_\ell+w_\nu+w_d.
\end{eqnarray}

Let us focus first on the driving superpotential $w_d$, which at LO in this model is given by
\begin{eqnarray}
\nonumber&&w_d=g_1\xi^0(\chi\phi)_{\mathbf{1}}+g_2(\rho^0(\chi\phi)_{\mathbf{3}'})_{\mathbf{1}}+M_{\chi}(\chi^0\chi)_{\mathbf{1}}+g_3(\chi^0(\phi\phi)_{\mathbf{5}_1})_{\mathbf{1}}+g_4(\chi^0(\phi\phi)_{\mathbf{5}_2})_{\mathbf{1}}\\
\label{eq:alignment}&&~~+h_1(\varphi^0\varphi)_{\mathbf{1}}\zeta+h_2(\varphi^0(\varphi\varphi)_{\mathbf{5}_1})_{\mathbf{1}}+h_3(\varphi^0(\varphi\varphi)_{\mathbf{5}_2})_{\mathbf{1}}.
\end{eqnarray}
The charge assignments reported in Table \ref{tab:field2} indicate that the minimization equations of the driving potential can be separated into two decoupled sets: one for the neutrino sector and one for the charged lepton sector. In the charged lepton sector, the minimum is determined by the following conditions:
\begin{eqnarray}
\nonumber&&\frac{\partial w_d}{\partial\xi^0}=g_1(\chi_1\phi_1+\chi_2\phi_5+\chi_3\phi_4+\chi_4\phi_3+\chi_5\phi_2)=0\\
\nonumber&&\frac{\partial w_d}{\partial\rho^0_1}=g_2(2\chi_2\phi_5-\chi_3\phi_4+\chi_4\phi_3-2\chi_5\phi_2)=0\\
\nonumber&&\frac{\partial w_d}{\partial\rho^0_2}=g_2(-\sqrt{3}\,\chi_1\phi_4+\sqrt{2}\,\chi_2\phi_3-\sqrt{2}\,\chi_3\phi_2+\sqrt{3}\,\chi_4\phi_1)=0\\
\nonumber&&\frac{\partial w_d}{\partial\rho^0_3}=g_2(\sqrt{3}\,\chi_1\phi_3-\sqrt{3}\,\chi_3\phi_1+\sqrt{2}\,\chi_4\phi_5-\sqrt{2}\,\chi_5\phi_4)=0\\
\nonumber&&\frac{\partial w_d}{\partial\chi^0_1}=M_{\chi}\chi_1+2g_3(\phi^2_1+\phi_2\phi_5-2\phi_3\phi_4)+2g_4(\phi^2_1-2\phi_2\phi_5+\phi_3\phi_4)=0\\
\nonumber&&\frac{\partial w_d}{\partial\chi^0_2}=M_{\chi}\chi_5+2g_3(\phi_1\phi_5+\sqrt{6}\,\phi_2\phi_4)+g_4(-4\phi_1\phi_5+\sqrt{6}\,\phi^2_3)=0\\
\nonumber&&\frac{\partial w_d}{\partial\chi^0_3}=M_{\chi}\chi_4+g_3(-4\phi_1\phi_4+\sqrt{6}\,\phi^2_5)+2g_4(\phi_1\phi_4+\sqrt{6}\,\phi_2\phi_3)=0\\
\nonumber&&\frac{\partial w_d}{\partial\chi^0_4}=M_{\chi}\chi_3+g_3(-4\phi_1\phi_3+\sqrt{6}\,\phi^2_2)+2g_4(\phi_1\phi_3+\sqrt{6}\,\phi_4\phi_5)=0\\
&&\frac{\partial w_d}{\partial\chi^0_5}=M_{\chi}\chi_2+2g_3(\phi_1\phi_2+\sqrt{6}\,\phi_3\phi_5)+g_4(-4\phi_1\phi_2+\sqrt{6}\,\phi^2_4)=0,
\end{eqnarray}
which result in the solution\footnote{There is another solution in which $\langle\chi\rangle=(0,v_{\chi},0,0,0)$, and $\langle\phi\rangle=(0,0,0,v_{\phi},0)$, with $v_{\chi}=-\sqrt{6}\,g_4v^2_{\phi}/M_{\chi}$ and $v_{\phi}$ undetermined. However, this solution can be obtained by acting on Eq.~(\ref{eq:lepton_alignment}) with the $A_5$ group element $T^3ST^2ST^3S$, and consequently it is an equivalent solution.}
\begin{equation}
\label{eq:lepton_alignment}\langle\chi\rangle=(0,0,0,0,v_{\chi}),~~~~\langle\phi\rangle=(0,0,v_{\phi},0,0),
\end{equation}
in which
\begin{equation}
\label{eq:vacuum}v_{\chi}=-\sqrt{6}\,g_4v^2_{\phi}/M_{\chi},~~~~v_{\phi}~\mathrm{undetermined}.
\end{equation}
Under the action of $T$, we have $T\langle\chi\rangle=\rho^4\langle\chi\rangle$ and $T\langle\phi\rangle=\rho^2\langle\phi\rangle$. As a result, the $Z_5$ symmetry is broken completely: no residual symmetry is preserved in the charged lepton sector. For the neutrino sector, the minimization equations take the following form:
\begin{eqnarray}
\nonumber&&\frac{\partial w_d}{\partial\varphi^{0}_1}=h_1\varphi_1\zeta+2h_2(\varphi^2_1+\varphi_2\varphi_5-2\varphi_3\varphi_4)+2h_3(\varphi^2_1-2\varphi_2\varphi_5+\varphi_3\varphi_4)=0\\
\nonumber&&\frac{\partial w_d}{\partial\varphi^{0}_2}=h_1\varphi_5\zeta+2h_2(\varphi_1\varphi_5+\sqrt{6}\,\varphi_2\varphi_4)+h_3(-4\varphi_1\varphi_5+\sqrt{6}\,\varphi^2_3)=0\\
\nonumber&&\frac{\partial w_d}{\partial\varphi^{0}_3}=h_1\varphi_4\zeta+h_2(-4\varphi_1\varphi_4+\sqrt{6}\,\varphi^2_5)+2h_3(\varphi_1\varphi_4+\sqrt{6}\,\varphi_2\varphi_3)=0\\
\nonumber&&\frac{\partial w_d}{\partial\varphi^{0}_4}=h_1\varphi_3\zeta+h_2(-4\varphi_1\varphi_3+\sqrt{6}\,\varphi^2_2)+2h_3(\varphi_1\varphi_3+\sqrt{6}\,\varphi_4\varphi_5)=0\\
&&\frac{\partial w_d}{\partial\varphi^{0}_5}=h_1\varphi_2\zeta+2h_2(\varphi_1\varphi_2+\sqrt{6}\varphi_3\varphi_5)+h_3(-4\varphi_1\varphi_2+\sqrt{6}\,\varphi^2_4)=0.
\end{eqnarray}
The minimum thus takes the form
\begin{equation}
\label{eq:neutrino_alignment}\langle\zeta\rangle=v_{\zeta},~~~\langle\varphi\rangle=(\sqrt{\frac{2}{3}}(v_2+v_3),v_2,v_3,v_3,v_2),
\end{equation}
in which
\begin{eqnarray}
\nonumber&&h_2v_2(-v^2_2+2v_2v_3+4v^2_3)=h_3v_3(-v^2_3+2v_2v_3+4v^2_2)\\
&&v_{\zeta}=-2\sqrt{\frac{2}{3}}\,\frac{h_2}{h_1}(v_2+4v_3)-\sqrt{\frac{2}{3}}\;\frac{h_3}{h_1}\big[\frac{3v^2_3}{v_2}-4(v_2+v_3)\big].
\end{eqnarray}
In this sector, $A_5$ is broken to its $K_4$ subgroup by the nonvanishing VEVs of $\zeta$ and $\varphi$. Given the symmetry of $w_d$, we can generate other minima of the scalar potential by  acting on Eq.~(\ref{eq:lepton_alignment}) and Eq.~(\ref{eq:neutrino_alignment}) with group elements. However, these new minima are related to the original minimum by field redefinitions, and hence are physically equivalent. As a result, there is no loss of generality in selecting Eq.~(\ref{eq:lepton_alignment}) and Eq.~(\ref{eq:neutrino_alignment}) as the local minimum.

We turn now to the Yukawa superpotential for the charged leptons, which in this model takes the form
\begin{eqnarray}
\nonumber&&w_{\ell}=\frac{y_{\tau}}{\Lambda}((E^cL)_{\mathbf{5}}\phi)_{\mathbf{1}}h_d+\frac{y_{\mu_1}}{\Lambda^2}(E^cL)_{\mathbf{1}}(\chi\chi)_{\mathbf{1}}h_d+\frac{y_{\mu_2}}{\Lambda^2}((E^cL)_{\mathbf{3}}(\chi\chi)_{\mathbf{3}})_{\mathbf{1}}h_d\\
&&~~~~+\frac{y_{\mu_3}}{\Lambda^2}((E^cL)_{\mathbf{5}}(\chi\chi)_{\mathbf{5}_1})_{\mathbf{1}}h_d+\frac{y_{\mu_4}}{\Lambda^2}((E^cL)_{\mathbf{5}}(\chi\chi)_{\mathbf{5}_2})_{\mathbf{1}}h_d+ \ldots,
\end{eqnarray}
in which we have neglected to write higher dimensional operators that will be addressed later in the paper. After electroweak breaking and flavor symmetry breaking via Eq.~(\ref{eq:lepton_alignment}), we see that the charged lepton mass matrix is diagonal:
\begin{equation}
m_{\ell}=\left(\begin{array}{ccc}0&0&0\\
0 & 6y_{\mu_3}v^2_{\chi}/\Lambda^2 & 0\\
0 & 0 & \sqrt{6}\,y_{\tau}v_{\phi}/\Lambda
\end{array}\right)v_d.
\end{equation}
Clearly, the electron is massless at leading order; it will be shown later that its mass arises at next to leading (NLO) order. As the tau and muon masses are suppressed respectively by $1/\Lambda$ and $1/\Lambda^2$, the observed mass hierarchies can be reproduced naturally if $v_{\phi}/\Lambda$ and $v_{\chi}/\Lambda$ are of order $\lambda^2_c$, in which $\lambda_c\approx0.22$ is the Cabibbo angle. We note that the charged lepton mass hierarchies can thus be determined by the flavor symmetry itself without invoking further Froggatt-Nielsen mechanisms or fine tuning the couplings. As is common in discrete flavor symmetry model building, we assume that all VEVs are approximately at the same scale.  Hence we take $v_{\phi}/\Lambda\sim v_{\chi}/\Lambda\sim v_{\zeta}/\Lambda\sim v_2/\Lambda\sim v_3/\Lambda\sim\lambda^2_c$ in the following.

In the neutrino sector, the superpotential $w_\nu$ is given as follows:
\begin{equation}
w_{\nu}=\frac{y_{\nu_1}}{\Lambda}(N^cL)_{\mathbf{1}}\zeta h_u+\frac{y_{\nu_2}}{\Lambda}((N^cL)_{\mathbf{5}}\varphi)_{\mathbf{1}}h_u+\frac{1}{2}M(N^cN^c)_{\mathbf{1}}+...
\end{equation}
where once again we have neglected at this stage to write higher-order contributions. Note that we can always choose the mass parameter $M$ to be real and positive by performing a global phase transformation on the right-handed neutrinos. Given the set of flavon VEVs of Eq.~(\ref{eq:neutrino_alignment}), the neutrino Dirac mass matrix is
\begin{eqnarray}
m_D=\left(\begin{array}{ccc}
y_{\nu_1}v_{\zeta}+2\sqrt{\frac{2}{3}}\,y_{\nu_2}(v_2+v_3)  &\!\!\!\!\!\! -\sqrt{3}\,y_{\nu_2}v_2  &  -\sqrt{3}\,y_{\nu_2}v_2\\
-\sqrt{3}\,y_{\nu_2}v_2  & \sqrt{6}\,y_{\nu_2}v_3  &\!\!\!\!\!\!  y_{\nu_1}v_{\zeta}-\sqrt{\frac{2}{3}}\,y_{\nu_2}(v_2+v_3)\\
-\sqrt{3}y_{\nu_2}v_2  &\!\!\! \!\!\! y_{\nu_1}v_{\zeta}-\sqrt{\frac{2}{3}}\,y_{\nu_2}(v_2+v_3) & \sqrt{6}\,y_{\nu_2}v_3
\end{array}\right)\frac{v_u}{\Lambda},
\end{eqnarray}
and right-handed Majorana mass matrix is
\begin{eqnarray}
m_M=\left(\begin{array}{ccc}M & 0 & 0\\
0 & 0 &  M\\
0 & M & 0
\end{array}\right).
\end{eqnarray}
Note that the three right-handed neutrinos are exactly degenerate. Integrating out the heavy degrees of freedom, we obtain the light neutrino mass matrix:
\begin{equation}
\label{eq:see-saw}m_{\nu}=-m^T_Dm^{-1}_Mm_D=U^{*}_{\nu}\mathrm{diag}(m_1,m_2,m_3)U^{\dagger}_{\nu}.
\end{equation}
In the above, the unitary matrix $U_{\nu}$ is given by
\begin{equation}
\label{eq:Unu}U_{\nu}=U_{{\rm GR}1}\,\mathrm{diag}(ie^{-i\alpha_1},ie^{-i\alpha_2},e^{-i\alpha_3}),
\end{equation}
in which $U_{{\rm GR}1}$ is given by Eq.~(\ref{mnspgen}) for the GR1 mixing case ($\cot\theta_{12}=\phi_g$), and the phases $\alpha_i$ are given by
\begin{equation}
\alpha_1=\mathrm{arg}(A+B+C),~~~\alpha_2=\mathrm{arg}(A+B-C),~~~\alpha_3=\mathrm{arg}(A-2B),
\end{equation}
with
\begin{equation}
A=y_{\nu_1}\frac{v_{\zeta}}{\Lambda},~~~~B=\frac{y_{\nu_2}}{\sqrt{6}}\frac{v_2+4v_3}{\Lambda},~~~~C=\frac{\sqrt{30}\,y_{\nu_2}}{2}\frac{v_2}{\Lambda}.
\end{equation}
The light neutrino masses $m_1$, $m_2$ and $m_3$ are thus
\begin{equation}
m_1=|A+B+C|^2\frac{v^2_u}{M},~~~m_2=|A+B-C|\frac{v^2_u}{M},~~~m_3=|A-2B|^2\frac{v^2_u}{M}.
\end{equation}
The neutrino masses in this model thus depend on three complex parameters $A$, $B$ and $C$ as well as the overall factor $v^2_u/M$.  Therefore, $m_{1,2,3}$ {\it a priori} are unrelated and any type of mass hierarchy can be accommodated. In particular, there is no general neutrino mass sum rule constraint in this case.
\subsubsection{NLO corrections}
In this section, we address the next-to-leading corrections to the results presented above for our $A_5\times Z_3\times Z_3$ model.  Let us begin with the driving superpotential. Taking into account the NLO operators,
\begin{equation}
w_d=w^0_d+\delta w_d,
\end{equation}
in which $w^0_d$ is the leading order contribution presented previously, and $\delta w_d$ is the most general  $A_5\times Z_3 \times Z_3$ invariant quartic polynomial that is linear in the driving fields.  More precisely,  $\delta w_d$ can be written as
\begin{equation}
\delta w_d=\frac{1}{\Lambda}\sum^{8}_{i=1}x_i{\cal I}^{\xi^0}_i+\frac{1}{\Lambda}\sum^{19}_{i=1}r_i{\cal I}^{\rho^0}_i+\frac{1}{\Lambda}\sum^{11}_{i=1}c_i{\cal I}^{\chi^0}_i,
\end{equation}
where the $O(1)$ coefficients $x_i$, $r_i$ and $c_i$ are unconstrained by the family symmetry, and ${\cal I}^{\xi^0}_i$, ${\cal I}^{\rho^0}_i$ and ${\cal I}^{\chi^0}_i$ represent a basis of independent quartic invariants, in which the ${\cal I}^{\xi^0}_i$ are
\begin{eqnarray}
\nonumber&&{\cal I}^{\xi^0}_1=\xi^0(\chi(\chi\chi)_{\mathbf{5}_1})_{\mathbf{1}},~~~{\cal I}^{\xi^0}_2=\xi^0(\chi(\chi\chi)_{\mathbf{5}_2})_{\mathbf{1}},~~~{\cal I}^{\xi^0}_3=\xi^0(\phi(\phi\phi)_{\mathbf{5}_1})_{\mathbf{1}},~~~{\cal I}^{\xi^0}_4=\xi^0(\phi(\phi\phi)_{\mathbf{5}_2})_{\mathbf{1}},\\
&&{\cal I}^{\xi^0}_5=\xi^0\zeta^3,~~~{\cal I}^{\xi^0}_6=\xi^0\zeta(\varphi\varphi)_{\mathbf{1}},~~~{\cal I}^{\xi^0}_7=\xi^0(\varphi(\varphi\varphi)_{\mathbf{5}_1})_{\mathbf{1}},~~~{\cal I}^{\xi^0}_8=\xi^0(\varphi(\varphi\varphi)_{\mathbf{5}_2})_{\mathbf{1}},
\end{eqnarray}
while the ${\cal I}^{\rho^0}_i$ take the form
\begin{eqnarray}
\nonumber&&{\cal I}^{\rho^0}_1=(\rho^0(\chi(\chi\chi)_{\mathbf{3}})_{\mathbf{3}'})_{\mathbf{1}},~~~{\cal I}^{\rho^0}_2=(\rho^0(\chi(\chi\chi)_{\mathbf{3}'})_{\mathbf{3}'})_{\mathbf{1}},~~~{\cal I}^{\rho^0}_3=(\rho^0(\chi(\chi\chi)_{\mathbf{4}_S})_{\mathbf{3}'})_{\mathbf{1}},\\
\nonumber&&{\cal I}^{\rho^0}_4=(\rho^0(\chi(\chi\chi)_{\mathbf{4}_A})_{\mathbf{3}'})_{\mathbf{1}},~~~{\cal I}^{\rho^0}_5=(\rho^0(\chi(\chi\chi)_{\mathbf{5}_1})_{\mathbf{3}'})_{\mathbf{1}},~~~{\cal I}^{\rho^0}_6=(\rho^0(\chi(\chi\chi)_{\mathbf{5}_2})_{\mathbf{3}'})_{\mathbf{1}},\\
\nonumber&&{\cal I}^{\rho^0}_7=(\rho^0(\phi(\phi\phi)_{\mathbf{3}})_{\mathbf{3}'})_{\mathbf{1}},~~~{\cal I}^{\rho^0}_8=(\rho^0(\phi(\phi\phi)_{\mathbf{3}'})_{\mathbf{3}'})_{\mathbf{1}},~~~{\cal I}^{\rho^0}_9=(\rho^0(\phi(\phi\phi)_{\mathbf{4}_S})_{\mathbf{3}'})_{\mathbf{1}},\\
\nonumber&&{\cal I}^{\rho^0}_{10}=(\rho^0(\phi(\phi\phi)_{\mathbf{4}_A})_{\mathbf{3}'})_{\mathbf{1}},~~~{\cal I}^{\rho^0}_{11}=(\rho^0(\phi(\phi\phi)_{\mathbf{5}_1})_{\mathbf{3}'})_{\mathbf{1}},~~~{\cal I}^{\rho^0}_{12}=(\rho^0(\phi(\phi\phi)_{\mathbf{5}_2})_{\mathbf{3}'})_{\mathbf{1}},\\
\nonumber&&{\cal I}^{\rho^0}_{13}=(\rho^0(\varphi\varphi)_{\mathbf{3}'})_{\mathbf{1}}\zeta,~~~{\cal I}^{\rho^0}_{14}=(\rho^0(\varphi(\varphi\varphi)_{\mathbf{3}})_{\mathbf{3}'})_{\mathbf{1}},~~~{\cal I}^{\rho^0}_{15}=(\rho^0(\varphi(\varphi\varphi)_{\mathbf{3}'})_{\mathbf{3}'})_{\mathbf{1}},\\
\nonumber&&{\cal I}^{\rho^0}_{16}=(\rho^0(\varphi(\varphi\varphi)_{\mathbf{4}_S})_{\mathbf{3}'})_{\mathbf{1}},~~~{\cal I}^{\rho^0}_{17}=(\rho^0(\varphi(\varphi\varphi)_{\mathbf{4}_A})_{\mathbf{3}'})_{\mathbf{1}},~~~{\cal I}^{\rho^0}_{18}=(\rho^0(\varphi(\varphi\varphi)_{\mathbf{5}_1})_{\mathbf{3}'})_{\mathbf{1}},\\
&&{\cal I}^{\rho^0}_{19}=(\rho^0(\varphi(\varphi\varphi)_{\mathbf{5}_2})_{\mathbf{3}'})_{\mathbf{1}},
\end{eqnarray}
and the ${\cal I}^{\chi^0}_i$ are given by
\begin{eqnarray}
\nonumber&&{\cal I}^{\chi^0}_1=(\chi^0\phi)_{\mathbf{1}}(\chi\chi)_{\mathbf{1}},~~~{\cal I}^{\chi^0}_2=(\chi^0(\phi(\chi\chi)_{\mathbf{3}})_{\mathbf{5}})_{\mathbf{1}},~~~{\cal I}^{\chi^0}_3=(\chi^0(\phi(\chi\chi)_{\mathbf{3}'})_{\mathbf{5}})_{\mathbf{1}},\\
\nonumber&&{\cal I}^{\chi^0}_4=(\chi^0(\phi(\chi\chi)_{\mathbf{4}_S})_{\mathbf{5}_1})_{\mathbf{1}},~~~{\cal I}^{\chi^0}_5=(\chi^0(\phi(\chi\chi)_{\mathbf{4}_S})_{\mathbf{5}_2})_{\mathbf{1}},~~~{\cal I}^{\chi^0}_6=(\chi^0(\phi(\chi\chi)_{\mathbf{4}_A})_{\mathbf{5}_1})_{\mathbf{1}},\\
\nonumber&&{\cal I}^{\chi^0}_7=(\chi^0(\phi(\chi\chi)_{\mathbf{4}_A})_{\mathbf{5}_2})_{\mathbf{1}},~~~{\cal I}^{\chi^0}_8=(\chi^0(\phi(\chi\chi)_{\mathbf{5}_1})_{\mathbf{5}_1})_{\mathbf{1}},~~~{\cal I}^{\chi^0}_9=(\chi^0(\phi(\chi\chi)_{\mathbf{5}_1})_{\mathbf{5}_2})_{\mathbf{1}},\\
&&{\cal I}^{\chi^0}_{10}=(\chi^0(\phi(\chi\chi)_{\mathbf{5}_2})_{\mathbf{5}_1})_{\mathbf{1}},~~~{\cal I}^{\chi^0}_{11}=(\chi^0(\phi(\chi\chi)_{\mathbf{5}_2})_{\mathbf{5}_2})_{\mathbf{1}}.
\end{eqnarray}
Note that due to the auxiliary $Z_3\times Z_3$ symmetry of our model, the subleading corrections to the driving superpotential that are proportional to $\varphi^0$ are suppressed by $1/\Lambda^2$. Therefore, the solutions for the minima of $\zeta$ and $\varphi$ given in  Eq.~(\ref{eq:neutrino_alignment}) is unchanged at this order and it is only modified at the next to next to leading order (NNLO). For the remaining fields, the new set of vacuum values is obtained as usual by solving the vanishing F-term constraints. In so doing, we seek a solution that perturbs the original set of minima to first order in $1/\Lambda$. After a lengthy calculation, we find that there are nontrivial shifts of $\phi$ and $\chi$:
\begin{equation}
\label{eq:shifts}\langle\chi\rangle=(\delta v_{\chi_1},0, \delta v_{\chi_3},\delta v_{\chi_4},v_{\chi}),~~~\langle\phi\rangle=(\delta v_{\phi_1},\delta v_{\phi_2},v_{\phi},\delta v_{\phi_4},0).
\end{equation}
Note that the VEVs of the second component of $\chi$ and the fifth component of $\phi$ still vanish. For the remaining fields, the shifts $\delta v_{\chi_1}$, $\delta v_{\chi_3}$, $\delta v_{\chi_4}$, $\delta v_{\phi_1}$, $\delta v_{\phi_2}$, and $\delta v_{\phi_4}$ are given as follows:
\begin{eqnarray}
\nonumber&&\delta v_{\chi_1}=\big[\sqrt{2}\,d_3g_2+2d_2(g_4-2g_3)\big]\frac{v^3_{\chi}}{4\sqrt{3}\,g_2g_3\Lambda v_{\phi}}\\
\nonumber&&\delta v_{\chi_3}=\frac{(-4g_3+2g_4)v_{\chi}}{\sqrt{6}\,g_4v_{\phi}}\delta v_{\phi_1}\\
\nonumber&&\delta v_{\chi_4}=-\frac{2d_1v^3_{\zeta}}{3g_1\Lambda v_{\phi}}\\
\nonumber&&\delta v_{\phi_2}=-\frac{d_1v^3_{\zeta}}{3g_1\Lambda v_{\chi}}\\
&&\delta v_{\phi_4}=\big(\sqrt{2}\,d_2g_4+d_3g_2\big)\frac{v^2_{\chi}}{4g_2g_3\Lambda},
\end{eqnarray}
in which the parameters $d_1$, $d_2$ and $d_3$ are
\begin{eqnarray}
\nonumber&&d_1=x_5+4x_6(2v^2_2+v_2v_3+2v^2_3)/(3v^2_{\zeta})+(2/3)^{3/2}x_7(v_2+4v_3)(11v^2_2-2v_2v_3-4v^2_3)/v^3_{\zeta}\\
\nonumber&&~~~+(2/3)^{3/2}x_8(4v_2+v_3)(11v^2_3-2v_2v_3-4v^2_2)/v^3_{\zeta}\\
\nonumber&&d_2=-2\sqrt{3}\,(2r_3+r_5)\\
&&d_3=-4\sqrt{6}\,c_4+16\sqrt{3}\,c_5-2\sqrt{6}\,c_8+\sqrt{6}\,c_9.
\end{eqnarray}
Parametrizing the ratio of a generic flavon VEV to the cutoff scale $\Lambda$ by the small quantity $\varepsilon$, we see that
$\delta v_{\chi_{1,4}}/v_{\chi}$ and $\delta v_{\phi_{2,4}}/v_{\phi}$ are both of $O(\varepsilon)$.  Furthermore, $\delta v_{\phi_1}$, which contributes to the mass of the electron, is undetermined at this order. To obtain a phenomenologically acceptable range of values for the electron mass, we therefore require that $\delta v_{\phi_1}$ is $O(\varepsilon^2v_{\phi})$. As a result of this choice, $\delta v_{\chi_3}$ is suppressed by $\varepsilon^2$ with respect to $v_{\chi}$ as well. This is clearly an assumption in this model, since in the absence of a theory for the higher-order terms, we have allowed for the most general higher-order corrections.  Indeed, it is a result that ultimately is due to our insistence on triplet embeddings of both the left-handed and right-handed SM leptons.  It is worth noting that this issue can in principle be overcome without resorting to triplet embeddings by extending the $A_5$ flavor symmetry to its double cover ${\cal I}'$, which has doublet representations in which the right-handed charged leptons can be embedded (note that ${\cal I}'$ has previously been used in this way for quark flavor model building \cite{Everett:2010rd}).

We turn now to the NLO corrections to the lepton masses and mixings.  In the charged lepton sector,  the following  operators can contribute:
\begin{eqnarray}
\nonumber&&(E^cL)_{\mathbf{1}}(\chi(\phi\phi)_{\mathbf{5}_1})_{\mathbf{1}}h_d/\Lambda^3,~~~(E^cL)_{\mathbf{1}}(\chi(\phi\phi)_{\mathbf{5}_2})_{\mathbf{1}}h_d/\Lambda^3,~~~((E^cL)_{\mathbf{3}}(\chi(\phi\phi)_{\mathbf{3}})_{\mathbf{3}})_{\mathbf{1}}h_d/\Lambda^3,\\
\nonumber&&((E^cL)_{\mathbf{3}}(\chi(\phi\phi)_{\mathbf{3}'})_{\mathbf{3}})_{\mathbf{1}}h_d/\Lambda^3,~~~((E^cL)_{\mathbf{3}}(\chi(\phi\phi)_{\mathbf{4}_S})_{\mathbf{3}})_{\mathbf{1}}h_d/\Lambda^3,~~~((E^cL)_{\mathbf{3}}(\chi(\phi\phi)_{\mathbf{4}_A})_{\mathbf{3}})_{\mathbf{1}}h_d/\Lambda^3,\\
\nonumber&&((E^cL)_{\mathbf{3}}(\chi(\phi\phi)_{\mathbf{5}_1})_{\mathbf{3}})_{\mathbf{1}}h_d/\Lambda^3,~~~((E^cL)_{\mathbf{3}}(\chi(\phi\phi)_{\mathbf{5}_2})_{\mathbf{3}})_{\mathbf{1}}h_d/\Lambda^3,~~~((E^cL)_{\mathbf{5}}\chi)_{\mathbf{1}}(\phi\phi)_{\mathbf{1}}h_d/\Lambda^3,\\
\nonumber&&((E^cL)_{\mathbf{5}}(\chi(\phi\phi)_{\mathbf{3}})_{\mathbf{5}})_{\mathbf{1}}h_d/\Lambda^3,~~~((E^cL)_{\mathbf{5}}(\chi(\phi\phi)_{\mathbf{3}'})_{\mathbf{5}})_{\mathbf{1}}h_d/\Lambda^3,~~~((E^cL)_{\mathbf{5}}(\chi(\phi\phi)_{\mathbf{4}_S})_{\mathbf{5}_1})_{\mathbf{1}}h_d/\Lambda^3,\\
\nonumber&&((E^cL)_{\mathbf{5}}(\chi(\phi\phi)_{\mathbf{4}_S})_{\mathbf{5}_2})_{\mathbf{1}}h_d/\Lambda^3,~~~((E^cL)_{\mathbf{5}}(\chi(\phi\phi)_{\mathbf{4}_A})_{\mathbf{5}_1})_{\mathbf{1}}h_d/\Lambda^3,~~~((E^cL)_{\mathbf{5}}(\chi(\phi\phi)_{\mathbf{4}_A})_{\mathbf{5}_2})_{\mathbf{1}}h_d/\Lambda^3,\\
\nonumber&&((E^cL)_{\mathbf{5}}(\chi(\phi\phi)_{\mathbf{5}_1})_{\mathbf{5}_1})_{\mathbf{1}}h_d/\Lambda^3,~~~((E^cL)_{\mathbf{5}}(\chi(\phi\phi)_{\mathbf{5}_1})_{\mathbf{5}_2})_{\mathbf{1}}h_d/\Lambda^3,~~~((E^cL)_{\mathbf{5}}(\chi(\phi\phi)_{\mathbf{5}_2})_{\mathbf{5}_1})_{\mathbf{1}}h_d/\Lambda^3,\\
&&((E^cL)_{\mathbf{5}}(\chi(\phi\phi)_{\mathbf{5}_2})_{\mathbf{5}_2})_{\mathbf{1}}h_d/\Lambda^3.
\end{eqnarray}
For the leading order minimum given in Eq.~(\ref{eq:lepton_alignment}), it is straightforward to see that the above higher dimensional operators only lead to corrections of $O(\varepsilon^3v_d)$ to $m_{{\ell}_{22}}$. Hence, the NLO corrections to the charged lepton mass matrix dominantly arise from the modified vacuum alignment. Inserting the shifted values of $\chi$ and $\phi$ given in Eq.~(\ref{eq:shifts}) into the leading order terms of $w_{\ell}$, we find that the charged lepton mass matrix can be parameterized as follows:
\begin{equation}
m_{\ell}=\left(\begin{array}{ccc}2a_1\varepsilon^2 & b_1\varepsilon^2 & b_2\varepsilon\\
b_1\varepsilon^2  & a_2\varepsilon & -a_1\varepsilon^2\\
b_2\varepsilon & -a_1\varepsilon^2  & a_3
\end{array}\right)\varepsilon v_d,
\end{equation}
in which the complex coefficients $a_i$ and $b_i$ have $O(1)$ magnitudes. Consequently, the unitary matrix $U_{\ell}$, which is the transformation of the charged leptons to the physical basis in which $m^{\dagger}_{\ell}m_{\ell}$ is diagonal, is of the form
\begin{equation}
U_{\ell}\simeq\left(\begin{array}{ccc}
1  & (\frac{b_1}{a_2}\varepsilon)^{*}  &  (\frac{b_2}{a_3}\varepsilon)^{*} \\
-\frac{b_1}{a_2}\varepsilon  &  1  &  -(\frac{a_1}{a_3}\varepsilon^2)^{*}  \\
-\frac{b_2}{a_3}\varepsilon   &  \frac{a_1}{a_3}\varepsilon^2  & 1
\end{array}\right),
\end{equation}
and the charged lepton masses are
\begin{equation}
m_e\simeq|(2a_1-b^2_2/a_3)\varepsilon^3v_d|,~~~m_{\mu}\simeq|a_2\varepsilon^2v_d|,~~~m_{\tau}\simeq|a_3\varepsilon v_d|.
\end{equation}
Hence, a realistic pattern of charged lepton masses can be produced. In the neutrino sector, the subleading operators are obtained by inserting the product $\chi\phi$ in all possible ways in the LO operators of $w_{\nu}$ and then extracting the $A_5$ invariants. As we have shown above, the leading order structure of the neutrino flavons remains intact at NLO, and hence the corrections to the neutrino mass matrix only arise at NNLO.  Therefore, the light neutrino mass matrix $m_{\nu}$ given in Eq.~(\ref{eq:see-saw}) and the diagonalization matrix $U_{\nu}$ of Eq.~(\ref{eq:Unu}) are unchanged at NLO. Taking into account the corrections from the charged lepton sector, the lepton mixing angles are modified as follows:
\begin{eqnarray}
\nonumber&&\sin\theta_{13}\simeq\left \vert \left (\frac{b_1}{a_2}-\frac{b_2}{a_3}\right )\frac{\varepsilon}{\sqrt{2}}\right \vert\\
\nonumber&&\sin^2\theta_{12}\simeq\frac{5-\sqrt{5}}{10}-\frac{1}{\sqrt{10}}\left (\frac{b_1}{a_2}+\frac{b_2}{a_3}+c.c. \right)\varepsilon\\
&&\sin^2\theta_{23}\simeq\frac{1}{2}+\frac{1}{4}\left [\left |\frac{b_1}{a_2}-\frac{b_2}{a_3}\right |^2-2\left (\frac{a_1}{a_3}+c.c.\right )\right ]\varepsilon^2.
\end{eqnarray}
We see that $\theta_{13}$ and $\theta_{12}$ both receive corrections of $O(\varepsilon)\sim\lambda^2_c$, while the deviation of the atmospheric angle $\theta_{23}$ from its GR1 mixing value of $\pi/4$ is only generated at $O(\varepsilon^2)$.

\begin{figure}[hptb]
\begin{center}
\begin{tabular}{cc}
\includegraphics[scale=.30]{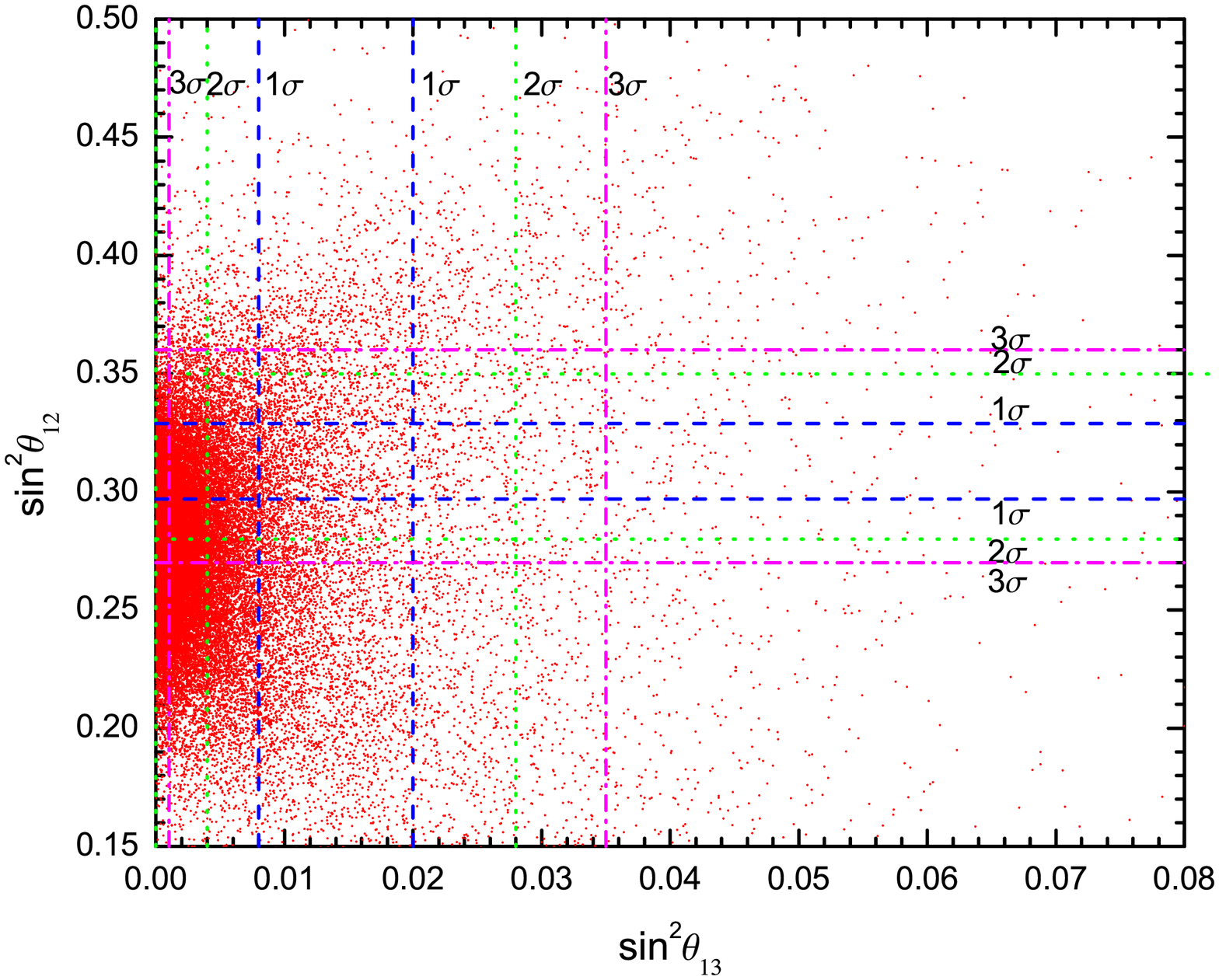} & \includegraphics[scale=.30]{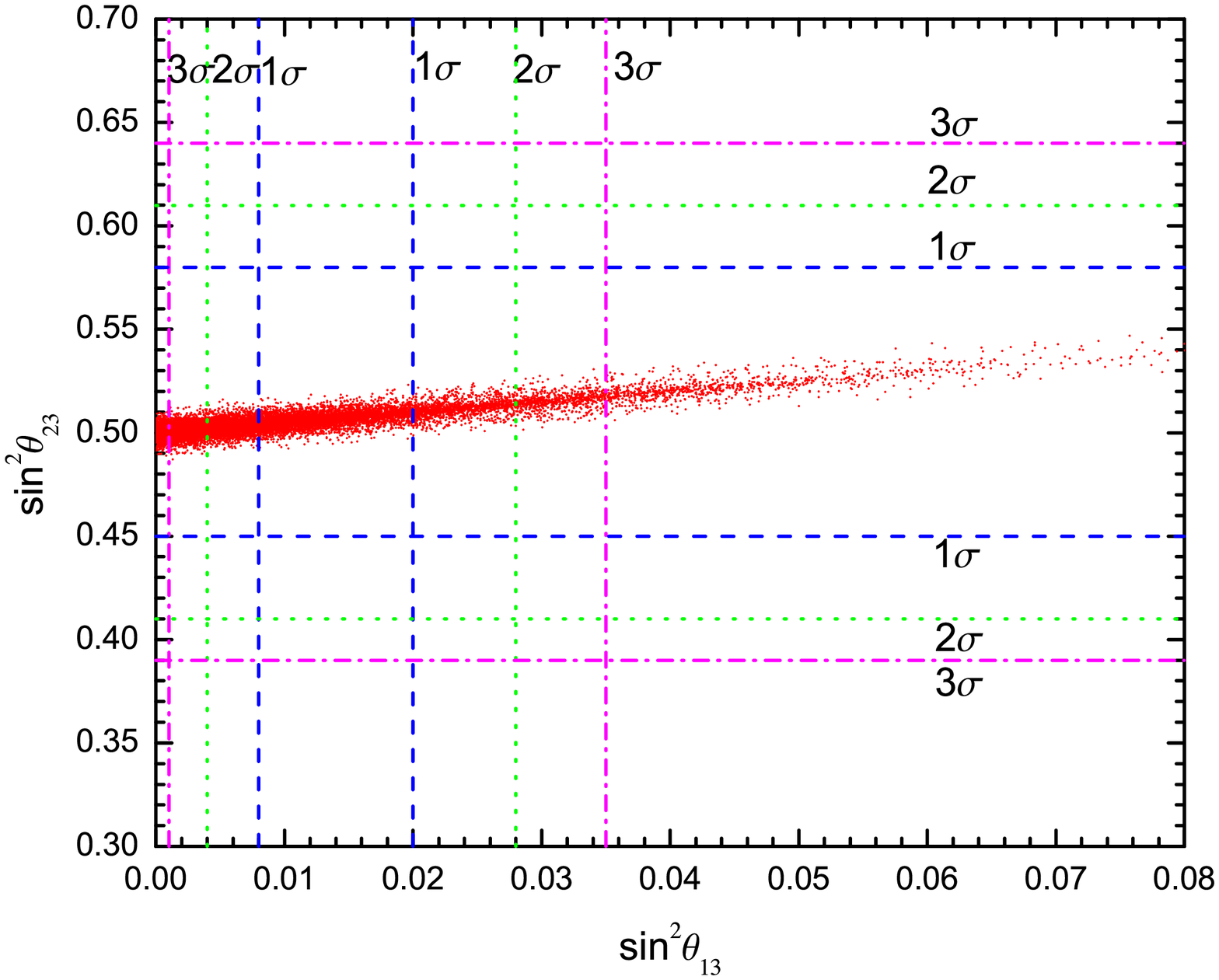}
\end{tabular}
\caption{\label{fig:mixing_angles}  $\sin^2\theta_{12}$ and $\sin^2\theta_{23}$ as a function of $\sin^2\theta_{13}$. The horizontal and vertical lines represent the $1\sigma$, $2\sigma$ and $3\sigma$ bounds on the mixing angles taken from Ref.\cite{Schwetz:2011zk}.}
\end{center}
\end{figure}

To investigate the predictions of our model in light of the current experimental ranges of the lepton mixing angles as described in the introduction and to cross-check the reliability of our analytical estimates, we have performed a numerical analysis in which the LO and NLO coefficients are taken to be random complex numbers with absolute value in the range of [1/3, 3], the corresponding phases are varied between 0 and $2\pi$, and $\varepsilon$ is set to 0.04 ($\sim \lambda_c^2$). The results of our analysis are shown in Fig.\ref{fig:mixing_angles}.   We see that a large set of points fall into the region in which  $\sin^2\theta_{13}$ and $\sin^2\theta_{12}$ are in the $3\sigma$ interval, while $\theta_{23}$ is in the $1\sigma$ range for almost all the points. However, it is clear from the parametric analysis given above that a value of $\theta_{13}$ near the present upper bound ({\it i.e.}, $\sim \lambda_c$) would be unnatural in this  model.

\section{Summary and Conclusions}
In this paper, we have studied the symmetry properties of lepton flavor mixing models in which at leading order the atmospheric mixing angle is $\pi/4$, the reactor angle $\theta_{13}$ is zero, and the solar mixing angle is related to the golden ratio, $\phi_g=(1+\sqrt{5})/2$. We considered both of the proposed relations available in the literature:  GR1 mixing, in which $\cot\theta_{12}=\phi_g$, and GR2 mixing, in which $\cos\theta_{12}=\phi_g/2$.  We first investigated the symmetry properties of the neutrino mass matrices as given in a basis in which the charged leptons are diagonal, and found that in each case there is a Klein four ($K_4\simeq Z_2\times Z_2$) symmetry that is preserved.  Next, we investigated the question of determining the minimal ({\it i.e.}, the smallest) discrete family symmetry group that encodes the symmetries of the fermion mass matrices.  We found that $A_5$ is the minimal group for GR1 mixing, and that up to a very high group order, there is no minimal discrete group that contains all of the symmetries needed for GR2 mixing.   If the $A_5$ family  symmetry is broken to $Z_5$ and $K_4$ in the charged lepton and neutrino sectors respectively, the GR1 mixing pattern follows immediately without fine-tuning.

We then focused our attention on models of GR1 mixing  based on $A_5$ with this pattern of flavor symmetry breaking, and constructed two prototype supersymmetric lepton flavor models in which the SM leptons are all assigned to triplet representations of $A_5$ and GR1 mixing is exactly reproduced at leading order.  In the first model (Model 1), the complete family symmetry is $A_5\times Z_5$, which is broken to $Z_5$ and $K_4$ as described.  The model predicts the neutrino mass sum rule $m_1^{-1}+m_2^{-1}=m_3^{-1}$, and the lightest neutrino mass is predicted to be larger than 0.011 eV (0.028 eV) for the normal (inverted) mass hierarchy.  Furthermore, the Majorana phases and $\vert m_{ee}\vert$ can also be expressed in terms of the light neutrino masses. However, although GR1 mixing is achieved without fine-tuning, the model has a drawback in that a separate fine-tuning of the charged lepton mass parameters is needed at leading order to achieve a phenomenologically acceptable mass hierarchy.

To improve upon this situation, we constructed a second model (Model 2), in which the family symmetry is $A_5\times Z_3\times Z_3$. In the neutrino sector, $A_5$ is again broken to $K_4$, and both the normal and inverted hierarchy patterns of the light neutrino masses can be accommodated. In the charged lepton sector, the $Z_5$ is further broken, and the charged lepton mass hierarchies are controlled by the symmetry breaking parameters $\varepsilon\sim \lambda_c^2$ such that there is no need to invoke a Froggatt-Nielsen $U(1)$ symmetry. In this model, the tau and muon masses are of $O(\varepsilon v_d)$ and $O(\varepsilon^2v_d)$ respectively, and electron is predicted to be massless at LO.  We next studied the NLO corrections due to higher dimensional operators are in detail, and fount that the neutrino sector remains intact to NLO, but the charged lepton mass matrix receives corrections that generate small off-diagonal elements as well as the electron mass.  To obtain an acceptable value for the electron mass, it is necessary to assume that the shift in the relevant flavon field value $\delta v_{\phi_1}/v_{\phi}$ is suppressed at the level of $O(\varepsilon^2)$ instead of $O(\varepsilon)$. The lepton mixing angles $\theta_{12}$ and $\theta_{13}$ are shifted from their GR1 values by $O(\varepsilon)\sim \lambda_c^2$ while $\theta_{23}$ receives corrections of $O(\varepsilon^2)\sim \lambda_c^4$.

This model is consistent with current lepton data, as we verified explicitly by a numerical analysis.  However, if the recent experimental hints of a potentially much larger value of $\theta_{13}$ are in fact verified in future experiments, this model would be strongly disfavored.   Indeed, this tension with a larger value of $\theta_{13}$ is a common feature within many lepton flavor models based on discrete family symmetries in which the reactor angle is zero at leading order (such as tri-bimaximal mixing models and others), as often the corrections  should at most be of $O(\lambda_c^2)$ to keep the solar mixing angle in a phenomenologically acceptable range. Hence,  a precise measurement of $\theta_{13}$ is a critically important test not only of this model, but also of many models within the general framework of discrete non-Abelian family symmetries.

\section*{Acknowledgements}

G.J.D is supported by the National Natural Science Foundation of China under Grant No~10905053, Chinese Academy KJCX2-YW-N29 and the 973 project with Grant No.~2009CB825200.  The work of L.E. and A.S. is supported by the U.~S.~Department of Energy under the contract DE-FG-02-95ER40896.

\appendix
\section*{Appendix: $A_5$ Representations and Clebsch-Gordan Coefficients}

As discussed in Section 2, we wish to work in a basis that satisfies Eq.~(\ref{eq:generator}), {\it i.e.}, one in which the order 5 generator $T$ is diagonal.  To determine this basis, we begin by noting that $A_5$ can be generated by two elements $X$ and $Y$ that satisfy the following relations \cite{A5everettstuart}:
\begin{equation}
X^2=Y^5=(Y^2XY^3XY^{-1}XYXY^{-1})^3=1.
\end{equation}
The explicit forms of these elements can be derived by carefully selecting several of the $A_5$ group representation matrices in the Shirai basis \cite{shirai}, which has been done in \cite{A5everettstuart}.  For completeness, we list the explicit representations below, beginning with the
 the generators of the three-dimensional irreducible representations.  For the $\mathbf{3}$, $X_{\mathbf{3}}$ and $Y_{\mathbf{3}}$
are given by
 \begin{eqnarray}
X_{\mathbf{3}}=\frac{1}{2}\left( \begin{array}{ccc}
 -1 &\phi_g& \frac{1}{\phi_g}\\
 \phi_g&\frac{1}{\phi_g}& 1\\
 \frac{1}{\phi_g}&1&-\phi_g
 \end{array} \right ), \;\;\;\;
Y_{\mathbf{3}}=\frac{1}{2}\left( \begin{array}{ccc}
 1&\phi_g&\frac{1}{\phi_g}\\
 -\phi_g&\frac{1}{\phi_g}& 1\\
\frac{1}{\phi_g}&-1&\phi_g
 \end{array} \right ),
\end{eqnarray}
where as usual $\phi_g=(1+\sqrt{5})/2$ is the golden ratio.  For the $\mathbf{3}^\prime$,  the generators take the forms:
\begin{eqnarray}
\label{stmat3p}
 X_{\mathbf{3}^\prime}=\frac{1}{2}\left(
\begin{array}{ccc}
 -\phi_g& \frac{1}{\phi_g}&1\\
 \frac{1}{\phi_g}&-1& \phi_g\\
 1&\phi_g&\frac{1}{\phi_g}
 \end{array}
\right ),\;\;\;\;
Y_{\mathbf{3}^\prime}=\frac{1}{2}\left( \begin{array}{ccc}
 -\phi_g&-\frac{1}{\phi_g}&1\\
 \frac{1}{\phi_g}&1& \phi_g\\
- 1&\phi_g&-\frac{1}{\phi_g}
 \end{array} \right ).
\end{eqnarray}
For the $\mathbf{4}$-dimensional representation, we have
\begin{eqnarray}
\label{stmat4}
X_{\mathbf{4}}=\frac{1}{4}\left(
\begin{array}{cccc}
 -1 & -1& -3& -\sqrt{5} \\
 -1 & 3& 1 & -\sqrt{5} \\
 -3 &1 & -1 & \sqrt{5} \\
 -\sqrt{5} & -\sqrt{5}& \sqrt{5}& -1
\end{array}
\right),\;\;\;
Y_4=\frac{1}{4}\left(
\begin{array}{cccc}
 -1 & 1& -3 & \sqrt{5}\\
 -1 & -3 & 1 &\sqrt{5} \\
  3 & 1 &1 & \sqrt{5} \\
 \sqrt{5}& -\sqrt{5}& -\sqrt{5}& -1
\end{array}\right),
\end{eqnarray}
and for the $\mathbf{5}$, the order two generator is
\begin{eqnarray}
\label{smat5}
X_{\mathbf{5}}=\frac{1}{2}\left(
\begin{array}{ccccc}
 \frac{1-3 \phi_g}{4}  & \frac{ \phi_g^2}{2} & -\frac{1}{2\phi_g^2}  & \frac{\sqrt{5}}{2} & \frac{\sqrt{3} }{4\phi_g} \\
 \frac{\phi_g^2}{2} & 1 & 1 & 0 & \frac{ \sqrt{3}}{2\phi_g}\\
 -\frac{1}{2\phi_g^2}  & 1& 0 & -1 & -\frac{\sqrt{3}\phi_g}{2} \\
 \frac{\sqrt{5}}{2} & 0 & -1 & 1 & -\frac{\sqrt{3}}{2} \\
 \frac{\sqrt{3}}{4\phi_g}   & \frac{\sqrt{3}}{2\phi_g}  & -\frac{ \sqrt{3}\phi_g}{2} &
-\frac{\sqrt{3}}{2} & \frac{3\phi_g-1}{4}
\end{array}\right),
\end{eqnarray}
and the order five generator is
\begin{eqnarray}
\label{tmat5}
Y_{\mathbf{5}}= \frac{1}{2}\left(
\begin{array}{ccccc}
  \frac{1-3 \phi_g}{4} &- \frac{\phi_g^2}{2}  & -\frac{1}{2\phi_g^2} & -\frac{\sqrt{5}}{2} & \frac{\sqrt{3}}{4\phi_g}
\\
 \frac{\phi_g^2}{2}  & -1 & 1 & 0 & \frac{\sqrt{3}}{2\phi_g} \\
 \frac{1}{2\phi_g^2}   & 1& 0 & -1 & \frac{\sqrt{3}\phi_g}{2}  \\
 -\frac{\sqrt{5}}{2} & 0 & 1 & 1 & \frac{\sqrt{3}}{2} \\
 \frac{ \sqrt{3}}{4\phi_g} & -\frac{\sqrt{3}}{2\phi_g} & -\frac{ \sqrt{3}\phi_g }{2} &
\frac{\sqrt{3}}{2} & \frac{3\phi_g-1}{4}
\end{array}\right).
\end{eqnarray}
We note that Eqs.~(\ref{smat5})--(\ref{tmat5}) correct minor typos found in \cite{shirai}.

With the preceding representation matrices, it is straightforward to derive the following relationship between the Shirai basis and the generators $S$ and $T$ in Eq.~(\ref{eq:generator}):
\begin{equation}\label{basisrelations}
S=XY^3XY^2X,~~~~T=Y^4 .
\end{equation}
Applying the above transformation to the Shirai basis given above results in a $T$ that is not diagonal but an $S$ that is. Since we wish to work in the diagonal $T$ basis, we diagonalize $T$ via a unitary transformation $U$, and hence
\begin{equation}
S=U^{\dagger}XY^3XY^2XU,~~~~T=U^{\dagger}Y^4U,
\end{equation}
For the $\mathbf{3}$, this results in
\begin{eqnarray}\label{gen3dim}
S_{\mathbf{3}}=
\frac{1}{\sqrt{5}}\left(
\begin{array}{ccc}
 1 & -\sqrt{2} & -\sqrt{2} \\
 -\sqrt{2} & -\phi_g  & \frac{1}{\phi_g } \\
 -\sqrt{2} & \frac{1}{\phi_g } & -\phi_g
\end{array}
\right),~~\;\;\;
T_{\mathbf{3}}=\left(
\begin{array}{ccc}
 1 & 0 & 0 \\
 0 & \rho  & 0 \\
 0 & 0 & \rho ^4
\end{array}
\right),
\end{eqnarray}
with
\begin{equation}
U_{\mathbf{3}}=\frac{1}{\sqrt{2}\,5^{1/4}}\left(
\begin{array}{ccc}
-\sqrt{\frac{2}{\phi_g}}
& -\sqrt{\phi_g} & -\sqrt{\phi_g} \\ 0 & i5^{1/4} & - i5^{1/4}\\
-\sqrt{2\phi_g}&  \frac{1}{\sqrt{\phi_g}} & \frac{1}{\sqrt{\phi_g}}
\end{array}
\right),
\end{equation}
where $\rho=e^{\frac{2\pi i}{5}}$.
The generators of the $K_4$ group given in Eq.~(\ref{7}) can be expressed in terms of $S_{\mathbf{3}}$ and $T_{\mathbf{3}}$ as follows: $G_1=S_{\mathbf{3}}$, $G_2=T_{\mathbf{3}}^3S_{\mathbf{3}}T_{\mathbf{3}}^2S_{\mathbf{3}}T_{\mathbf{3}}^3$ and $G_3=T_{\mathbf{3}}^3S_{\mathbf{3}}T_{\mathbf{3}}^2S_{\mathbf{3}}T^3_{\mathbf{3}}S_{\mathbf{3}}$.  For the $\mathbf{3}'$, we have
\begin{eqnarray}\label{gen3pdim}
S_{\mathbf{3}'}=\frac{1}{\sqrt{5}}\left(
\begin{array}{ccc}
 -1 & \sqrt{2} & \sqrt{2} \\
 \sqrt{2} & -\frac{1}{\phi_g } & \phi_g  \\
 \sqrt{2} & \phi_g  & -\frac{1}{\phi_g }
\end{array}
\right),~~\;\;\;
T_{\mathbf{3}'}=\left(
\begin{array}{ccc}
 1 & 0 & 0 \\
 0 & \rho ^2 & 0 \\
 0 & 0 & \rho ^3
\end{array}
\right),
\end{eqnarray}
and
\begin{equation}
U_{\mathbf{3}'}=\frac{1}{\sqrt{2} \, 5^{1/4}}\left(
\begin{array}{ccc}
 0 & i 5^{1/4} & -i 5^{1/4} \\
  \sqrt{2\phi_g } & -\frac{1}{\sqrt{\phi_g }} & -\frac{1}{\sqrt{\phi_g }} \\
 \sqrt{\frac{2}{\phi_g}} & \sqrt{\phi_g } & \sqrt{\phi_g }
\end{array}
\right).
\end{equation}
This also contains $K_4$, in which $G_2=S_{\mathbf{3}'}$, $G_1=T_{\mathbf{3}'}^3S_{\mathbf{3}'}T_{\mathbf{3}'}^2S_{\mathbf{3}'}T_{\mathbf{3}'}^3$ and $G_3=T_{\mathbf{3}'}^3S_{\mathbf{3}'}T_{\mathbf{3}'}^2S_{\mathbf{3}'}T_{\mathbf{3}'}^3S_{\mathbf{3}'}$. Therefore, both the $3$ and $3^{\prime}$ irreducible representations in this basis for $A_5$ contain the matrices which generate the full set of  symmetries of the neutrino mass matrix in the case of GR1 mixing.  Turning now to the $\mathbf{4}$, we have
\begin{eqnarray}
S_{\mathbf{4}}=\frac{1}{\sqrt{5}}\left(
\begin{array}{cccc}
 1 & \frac{1}{\phi_g } & \phi_g  & -1 \\
 \frac{1}{\phi_g } & -1 & 1 & \phi_g  \\
 \phi_g  & 1 & -1 & \frac{1}{\phi_g } \\
 -1 & \phi_g  & \frac{1}{\phi_g } & 1
\end{array}
\right),~~\;\;\;
T_{\mathbf{4}}=\left(
\begin{array}{cccc}
 \rho  & 0 & 0 & 0 \\
 0 & \rho ^2 & 0 & 0 \\
 0 & 0 & \rho ^3 & 0 \\
 0 & 0 & 0 & \rho ^4
\end{array}
\right),
\end{eqnarray}
and
\begin{equation}
U_{\mathbf{4}}=\frac{1}{2} \left(
\begin{array}{cccc}
 1 & -1 & -1 & 1 \\
 \frac{i}{5^{1/4} \phi_g ^{3/2}} & \frac{i \phi_g ^{3/2}}{5^{1/4}} & -\frac{i \phi_g ^{3/2}}{5^{1/4}} & -\frac{i}{5^{1/4} \phi_g
   ^{3/2}} \\
 \frac{i \phi_g ^{3/2}}{5^{1/4}} & -\frac{i}{5^{1/4} \phi_g ^{3/2}} & \frac{i}{5^{1/4} \phi_g ^{3/2}} & -\frac{i \phi_g
   ^{3/2}}{5^{1/4}} \\
 1 & 1 & 1 & 1
\end{array}
\right).
\end{equation}
Finally, in the case of the $\mathbf{5}$,
\begin{eqnarray}
S_{\mathbf{5}}=\frac{1}{5} \left(
\begin{array}{ccccc}
 -1 & \sqrt{6} & \sqrt{6} & \sqrt{6} & \sqrt{6} \\
 \sqrt{6} & \frac{1}{\phi_g ^2} & -2 \phi_g  & \frac{2}{\phi_g } & \phi_g ^2 \\
 \sqrt{6} & -2 \phi_g  & \phi_g ^2 & \frac{1}{\phi_g ^2} & \frac{2}{\phi_g } \\
 \sqrt{6} & \frac{2}{\phi_g } & \frac{1}{\phi_g ^2} & \phi_g ^2 & -2 \phi_g  \\
 \sqrt{6} & \phi_g ^2 & \frac{2}{\phi_g } & -2 \phi_g  & \frac{1}{\phi_g ^2}
\end{array}
\right),~~\;\;\;
T_{\mathbf{5}}=\left(
\begin{array}{ccccc}
 1 & 0 & 0 & 0 & 0 \\
 0 & \rho  & 0 & 0 & 0 \\
 0 & 0 & \rho ^2 & 0 & 0 \\
 0 & 0 & 0 & \rho ^3 & 0 \\
 0 & 0 & 0 & 0 & \rho ^4
\end{array}
\right),
\end{eqnarray}
and
\begin{equation}
U_{\mathbf{5}}=\frac{1}{\sqrt{10}}\left(
\begin{array}{ccccc}
 -\frac{1}{\phi_g }\sqrt{\frac{3}{2}} & 1 & -\frac{1}{2} \sqrt{\phi_g ^4+8} & -\frac{1}{2} \sqrt{\phi_g ^4+8} & 1 \\
 0 & \frac{i 5^{1/4}}{\sqrt{\phi_g }} & -i 5^{1/4} \sqrt{\phi_g } & i 5^{1/4} \sqrt{\phi_g } & -\frac{i 5^{1/4}}{\sqrt{\phi_g }} \\
 0 & i 5^{1/4} \sqrt{\phi_g } & \frac{i 5^{1/4}}{\sqrt{\phi_g }} & -\frac{i 5^{1/4}}{\sqrt{\phi_g }} & -i 5^{1/4} \sqrt{\phi_g } \\
 \sqrt{6} & -1 & -1 & -1 & -1 \\
 \frac{\phi_g ^2}{\sqrt{2}} & \sqrt{3} & \frac{\sqrt{3}}{2 \phi_g } & \frac{\sqrt{3}}{2 \phi_g } & \sqrt{3}
\end{array}
\right)
\end{equation}
From the above representation matrices, it is straightforward to calculate the Clebsch-Gordan coefficients for the decomposition of the product representations, which we now report for this basis in detail.  We will use $\alpha_i$ to denote the elements of the first representation  and $\beta_i$ to indicate those of the second representation of the product, and we use the subscripts ``$A$" and ``$S$" to indicate a representation which is antisymmetric or symmetric, respectively.\\

\fbox{$\mathbf{3}\otimes\mathbf{3}=\mathbf{1}_S\oplus\mathbf{3}_A \oplus\mathbf{5}_S$}

\begin{equation}
\nonumber\mathbf{1}_S\sim \alpha _1 \beta _1+\alpha _2 \beta _3+\alpha _3 \beta _2
\end{equation}

\begin{equation}
\nonumber\mathbf{3}_A\sim\left(
\begin{array}{c}
 \alpha _2 \beta _3-\alpha _3 \beta _2 \\
 \alpha _1 \beta _2-\alpha _2 \beta _1 \\
 \alpha _3 \beta _1-\alpha _1 \beta _3
\end{array}
\right)
\end{equation}

\begin{equation}
\nonumber\mathbf{5}_S\sim\left(
\begin{array}{c}
 2 \alpha _1 \beta _1-\alpha _2 \beta _3-\alpha _3 \beta _2 \\
 -\sqrt{3} \alpha _1 \beta _2-\sqrt{3} \alpha _2 \beta _1 \\
 \sqrt{6} \alpha _2 \beta _2 \\
 \sqrt{6} \alpha _3 \beta _3 \\
-\sqrt{3} \alpha _1 \beta _3-\sqrt{3} \alpha _3 \beta _1
\end{array}
\right)
\end{equation}

\fbox{$\mathbf{3}'\otimes\mathbf{3}'=\mathbf{1}_S\oplus\mathbf{3}'_A\oplus\mathbf{5}_S$}

\begin{equation}
\nonumber\mathbf{1}_S\sim \alpha _1 \beta _1+\alpha _2 \beta _3+\alpha _3 \beta _2
\end{equation}

\begin{equation}
\nonumber\mathbf{3}'_A\sim\left(
\begin{array}{c}
 \alpha _2 \beta _3-\alpha _3 \beta _2 \\
 \alpha _1 \beta _2-\alpha _2 \beta _1 \\
 \alpha _3 \beta _1-\alpha _1 \beta _3
\end{array}
\right)
\end{equation}

\begin{equation}
\nonumber\mathbf{5}_S\sim\left(
\begin{array}{c}
 2 \alpha _1 \beta _1-\alpha _2 \beta _3 -\alpha _3 \beta _2\\
 \sqrt{6} \alpha _3 \beta _3 \\
 -\sqrt{3} \alpha _1 \beta _2 -\sqrt{3} \alpha _2 \beta _1\\
 -\sqrt{3} \alpha _1 \beta _3-\sqrt{3} \alpha _3 \beta _1 \\
 \sqrt{6} \alpha _2 \beta _2
\end{array}
\right)
\end{equation}

\fbox{$\mathbf{3}\otimes\mathbf{3}'=\mathbf{4}\oplus\mathbf{5}$}

\begin{equation}
\nonumber\mathbf{4}\sim\left(
\begin{array}{c}
 \sqrt{2} \alpha _2 \beta _1+\alpha _3 \beta _2 \\
 -\sqrt{2} \alpha _1 \beta _2-\alpha _3 \beta _3 \\
 -\sqrt{2} \alpha _1 \beta _3-\alpha _2 \beta _2 \\
 \sqrt{2} \alpha _3 \beta _1+\alpha _2 \beta _3
\end{array}
\right)
\end{equation}

\begin{equation}
\nonumber\mathbf{5}\sim\left(
\begin{array}{c}
 \sqrt{3} \alpha _1 \beta _1 \\
 \alpha _2 \beta _1-\sqrt{2} \alpha _3 \beta _2 \\
 \alpha _1 \beta _2-\sqrt{2} \alpha _3 \beta _3 \\
 \alpha _1 \beta _3-\sqrt{2} \alpha _2 \beta _2 \\
 \alpha _3 \beta _1-\sqrt{2} \alpha _2 \beta _3
\end{array}
\right)
\end{equation}

\fbox{$\mathbf{3}\otimes\mathbf{4}=\mathbf{3}'\oplus\mathbf{4}\oplus\mathbf{5}$}

\begin{equation}
\nonumber\mathbf{3}'\sim\left(
\begin{array}{c}
 -\sqrt{2} \alpha _2 \beta _4-\sqrt{2} \alpha _3 \beta _1 \\
 \sqrt{2} \alpha _1 \beta _2-\alpha _2 \beta _1+\alpha _3 \beta _3 \\
 \sqrt{2} \alpha _1 \beta _3+\alpha _2 \beta _2-\alpha _3 \beta _4
\end{array}
\right)
\end{equation}

\begin{equation}
\nonumber\mathbf{4}\sim\left(
\begin{array}{c}
  \alpha _1 \beta _1-\sqrt{2}\alpha _3 \beta _2 \\
 -\alpha _1 \beta _2-\sqrt{2}\alpha _2 \beta _1 \\
  \alpha _1 \beta _3+\sqrt{2}\alpha _3 \beta _4 \\
  -\alpha _1 \beta _4+\sqrt{2}\alpha _2 \beta _3
\end{array}
\right)
\end{equation}

\begin{equation}
\nonumber\mathbf{5}\sim\left(
\begin{array}{c}
 \sqrt{6} \alpha _2
   \beta _4-\sqrt{6} \alpha _3 \beta _1 \\
 2\sqrt{2} \alpha _1 \beta _1+2 \alpha _3 \beta_2\\
 -\sqrt{2} \alpha _1 \beta_2+\alpha _2 \beta _1+3\alpha _3 \beta _3 \\
\sqrt{2} \alpha _1 \beta_3-3\alpha _2 \beta _2-\alpha _3 \beta _4\\
 -2\sqrt{2} \alpha _1 \beta _4-2 \alpha _2 \beta _3
\end{array}
\right)
\end{equation}

\fbox{$\mathbf{3}'\otimes\mathbf{4}=\mathbf{3}\oplus\mathbf{4}\oplus\mathbf{5}$}

\begin{equation}
\nonumber\mathbf{3}\sim\left(
\begin{array}{c}
 -\sqrt{2} \alpha _2 \beta _3-\sqrt{2} \alpha _3 \beta _2 \\
 \sqrt{2} \alpha _1 \beta _1+\alpha _2 \beta _4-\alpha _3 \beta _3 \\
  \sqrt{2} \alpha _1 \beta _4 -\alpha _2 \beta _2+\alpha _3 \beta _1
\end{array}
\right)
\end{equation}

\begin{equation}
\nonumber\mathbf{4}\sim\left(
\begin{array}{c}
 \alpha _1 \beta _1+\sqrt{2}\alpha _3 \beta _3 \\
  \alpha _1 \beta _2-\sqrt{2}\alpha _3 \beta _4 \\
 -\alpha _1 \beta _3+\sqrt{2}\alpha _2 \beta _1 \\
   -\alpha _1 \beta _4-\sqrt{2}\alpha _2 \beta _2
\end{array}
\right)
\end{equation}

\begin{equation}
\nonumber\mathbf{5}\sim\left(
\begin{array}{c}
 \sqrt{6} \alpha _2 \beta _3-\sqrt{6} \alpha _3
   \beta _2 \\
   \sqrt{2} \alpha _1 \beta _1-3\alpha _2 \beta _4-\alpha _3 \beta_3 \\
   2\sqrt{2} \alpha _1 \beta _2+2 \alpha _3 \beta _4
   \\
   -2\sqrt{2} \alpha _1 \beta_3-2\alpha _2 \beta _1 \\
 -\sqrt{2}\alpha _1 \beta _4+\alpha _2 \beta _2+3\alpha _3 \beta _1
\end{array}
\right)
\end{equation}

\fbox{$\mathbf{3}\otimes\mathbf{5}=\mathbf{3}\oplus\mathbf{3}'\oplus\mathbf{4}\oplus\mathbf{5}$}

\begin{equation}
\nonumber\mathbf{3}\sim\left(
\begin{array}{c}
 -2 \alpha _1 \beta _1+\sqrt{3}\alpha _2 \beta _5+\sqrt{3}\alpha _3 \beta
   _2 \\
 \sqrt{3}\alpha _1 \beta
   _2+\alpha _2 \beta _1-\sqrt{6}\alpha _3 \beta _3 \\
 \sqrt{3}\alpha
   _1 \beta _5-\sqrt{6}\alpha _2 \beta _4+\alpha _3 \beta _1
\end{array}
\right)
\end{equation}

\begin{equation}
\nonumber\mathbf{3}'\sim\left(
\begin{array}{c}
 \sqrt{3} \alpha _1 \beta _1+\alpha _2 \beta _5+\alpha _3 \beta
   _2 \\
  \alpha _1 \beta _3-\sqrt{2}\alpha _2 \beta _2-\sqrt{2}\alpha _3 \beta
   _4 \\
 \alpha _1 \beta _4-\sqrt{2}\alpha _2 \beta _3-\sqrt{2}\alpha _3 \beta
   _5
\end{array}
\right)
\end{equation}

\begin{equation}
\nonumber\mathbf{4}\sim\left(
\begin{array}{c}
 2\sqrt{2} \alpha _1 \beta
   _2-\sqrt{6} \alpha _2 \beta _1+\alpha _3 \beta _3 \\
 -\sqrt{2}\alpha _1 \beta _3+2\alpha _2 \beta _2-3 \alpha
   _3 \beta _4 \\
 \sqrt{2}\alpha _1 \beta
   _4+3\alpha _2 \beta _3-2\alpha _3 \beta _5 \\
 -2\sqrt{2} \alpha _1 \beta _5-\alpha _2 \beta
   _4+\sqrt{6} \alpha _3 \beta _1
\end{array}
\right)
\end{equation}

\begin{equation}
\nonumber\mathbf{5}\sim\left(
\begin{array}{c}
 \sqrt{3} \alpha _2 \beta _5-\sqrt{3} \alpha _3
   \beta _2 \\
 -\alpha _1 \beta_2-\sqrt{3} \alpha _2 \beta _1-\sqrt{2}\alpha _3 \beta _3 \\
 -2 \alpha _1 \beta _3-\sqrt{2}\alpha _2 \beta _2 \\
 2\alpha _1 \beta _4+\sqrt{2}\alpha _3 \beta _5 \\
 \alpha_1 \beta _5+\sqrt{2}\alpha _2 \beta _4+ \sqrt{3} \alpha _3 \beta _1
\end{array}
\right)
\end{equation}

\fbox{$\mathbf{3}'\otimes\mathbf{5}=\mathbf{3}\oplus\mathbf{3}'\oplus\mathbf{4}\oplus\mathbf{5}$}

\begin{equation}
\nonumber\mathbf{3}\sim\left(
\begin{array}{c}
 \sqrt{3} \alpha _1 \beta
   _1+\alpha _2\beta_4+\alpha _3 \beta _3 \\
 \alpha _1 \beta _2-\sqrt{2}
   \alpha _2 \beta _5 -\sqrt{2}
   \alpha _3 \beta _4\\
 \alpha _1 \beta _5-\sqrt{2} \alpha_2 \beta_3-\sqrt{2} \alpha _3 \beta
   _2
\end{array}
\right)
\end{equation}

\begin{equation}
\nonumber\mathbf{3}'\sim\left(
\begin{array}{c}
 -2 \alpha _1 \beta _1+\sqrt{3}
   \alpha _2 \beta _4 +\sqrt{3}
   \alpha _3 \beta _3\\
 \sqrt{3}
   \alpha _1 \beta _3+\alpha _2 \beta _1-\sqrt{6}
   \alpha _3 \beta _5 \\
 \sqrt{3}
   \alpha _1 \beta _4-\sqrt{6}
   \alpha _2 \beta _2+\alpha _3 \beta _1
\end{array}
\right)
\end{equation}

\begin{equation}
\nonumber\mathbf{4}\sim\left(
\begin{array}{c}
 \sqrt{2} \alpha _1 \beta _2+3 \alpha _2
   \beta _5-2\alpha _3 \beta _4 \\
 2\sqrt{2} \alpha _1 \beta_3-\sqrt{6} \alpha _2 \beta _1+\alpha _3 \beta _5 \\
 -2\sqrt{2} \alpha _1 \beta _4-\alpha _2 \beta _2 +\sqrt{6} \alpha _3 \beta
   _1\\
 -\sqrt{2} \alpha _1\beta _5+2 \alpha _2
   \beta _3-3 \alpha _3 \beta _2
\end{array}
\right)
\end{equation}

\begin{equation}
\nonumber\mathbf{5}\sim\left(
\begin{array}{c}
 \sqrt{3} \alpha _2 \beta
   _4-\sqrt{3} \alpha _3 \beta _3
   \\
 2 \alpha _1 \beta _2+\sqrt{2}
   \alpha _3 \beta _4 \\
 -\alpha _1 \beta _3-\sqrt{3} \alpha _2 \beta
   _1-\sqrt{2}\alpha _3 \beta _5 \\
\alpha _1 \beta _4+\sqrt{2} \alpha _2 \beta_2 + \sqrt{3} \alpha _3 \beta
   _1\\
 -2\alpha _1 \beta _5-\sqrt{2} \alpha _2 \beta _3
\end{array}
\right)
\end{equation}

\fbox{$\mathbf{4}\otimes\mathbf{4}=\mathbf{1}_S\oplus\mathbf{3}_A\oplus\mathbf{3}'_A\oplus\mathbf{4}_S\oplus\mathbf{5}_S$}

\begin{equation}
\nonumber\mathbf{1}_S\sim\alpha _1\beta _4+\alpha _2 \beta _3+\alpha _3 \beta_2+\alpha _4 \beta _1
\end{equation}

\begin{equation}
\nonumber\mathbf{3}_A\sim\left(
\begin{array}{c}
 -\alpha _1 \beta _4+\alpha _2\beta_3-\alpha _3
   \beta _2+\alpha _4 \beta _1\\
 \sqrt{2} \alpha _2 \beta
   _4-\sqrt{2} \alpha _4 \beta _2
   \\
 \sqrt{2} \alpha _1 \beta
   _3-\sqrt{2} \alpha _3 \beta _1
\end{array}
\right)
\end{equation}

\begin{equation}
\nonumber\mathbf{3}'_A\sim\left(
\begin{array}{c}
\alpha _1 \beta _4 +\alpha _2 \beta _3-\alpha _3
   \beta _2 -\alpha _4 \beta _1\\
 \sqrt{2} \alpha _3 \beta
   _4-\sqrt{2} \alpha _4 \beta _3
   \\
 \sqrt{2} \alpha _1 \beta
   _2-\sqrt{2} \alpha _2 \beta _1
\end{array}
\right)
\end{equation}

\begin{equation}
\nonumber\mathbf{4}_S\sim\left(
\begin{array}{c}
 \alpha _2 \beta _4+\alpha _3\beta _3+\alpha _4 \beta _2 \\
 \alpha _1 \beta _1+\alpha _3 \beta _4 +\alpha _4 \beta _3\\
 \alpha _1\beta _2+\alpha _2 \beta _1+\alpha _4 \beta _4 \\
 \alpha _1 \beta _3+\alpha _2\beta _2+\alpha _3 \beta _1
\end{array}
\right)
\end{equation}

\begin{equation}
\nonumber\mathbf{5}_S\sim\left(
\begin{array}{c}
 \sqrt{3} \alpha _1 \beta _4-\sqrt{3} \alpha _2 \beta
   _3-\sqrt{3} \alpha _3 \beta_2+\sqrt{3} \alpha _4 \beta_1
   \\
 -\sqrt{2} \alpha _2 \beta _4+2
   \sqrt{2} \alpha _3 \beta_3-\sqrt{2} \alpha _4 \beta _2
   \\
 -2 \sqrt{2} \alpha _1 \beta
   _1+\sqrt{2} \alpha _3 \beta _4+\sqrt{2} \alpha _4 \beta
   _3
   \\
 \sqrt{2} \alpha _1 \beta
   _2+\sqrt{2} \alpha _2 \beta
   _1-2 \sqrt{2} \alpha _4 \beta
   _4 \\
 -\sqrt{2} \alpha _1 \beta _3+2\sqrt{2} \alpha _2 \beta_2-\sqrt{2} \alpha _3 \beta _1
\end{array}
\right)
\end{equation}

\fbox{$\mathbf{4}\otimes\mathbf{5}=\mathbf{3}\oplus\mathbf{3}'\oplus\mathbf{4}\oplus\mathbf{5}_1\oplus\mathbf{5}_2$}

\begin{equation}
\nonumber\mathbf{3}\sim\left(
\begin{array}{c}
 2 \sqrt{2} \alpha _1\beta_5-\sqrt{2} \alpha _2 \beta_4+\sqrt{2} \alpha _3 \beta_3-2 \sqrt{2} \alpha _4 \beta_2\\
 -\sqrt{6} \alpha _1 \beta_1+2 \alpha _2 \beta_5+3 \alpha_3 \beta _4-\alpha _4 \beta _3 \\
 \alpha _1 \beta _4-3 \alpha _2\beta _3-2\alpha _3 \beta _2+\sqrt{6} \alpha _4 \beta _1
\end{array}
\right)
\end{equation}

\begin{equation}
\nonumber\mathbf{3}'\sim\left(
\begin{array}{c}
 \sqrt{2} \alpha _1 \beta _5+2\sqrt{2} \alpha _2 \beta_4-2\sqrt{2} \alpha _3 \beta _3-\sqrt{2} \alpha _4 \beta _2   \\
 3\alpha _1 \beta _2-\sqrt{6} \alpha _2 \beta _1-\alpha _3 \beta _5+2 \alpha _4\beta _4 \\
 -2 \alpha_1 \beta _3+\alpha _2 \beta _2+\sqrt{6} \alpha _3 \beta_1-3 \alpha _4 \beta
   _5
\end{array}
\right)
\end{equation}

\begin{equation}
\nonumber\mathbf{4}\sim\left(
\begin{array}{c}
 \sqrt{3} \alpha _1 \beta _1-\sqrt{2} \alpha _2 \beta _5+\sqrt{2} \alpha _3 \beta_4-2\sqrt{2} \alpha _4 \beta_3
   \\
 -\sqrt{2} \alpha _1 \beta_2-\sqrt{3} \alpha _2 \beta_1+2 \sqrt{2} \alpha _3 \beta_5+\sqrt{2} \alpha _4 \beta_4 \\
 \sqrt{2} \alpha _1 \beta_3+2\sqrt{2} \alpha _2 \beta_2-\sqrt{3} \alpha _3 \beta _1-\sqrt{2} \alpha _4 \beta _5
   \\
 -2 \sqrt{2} \alpha _1 \beta_4+\sqrt{2} \alpha _2 \beta_3-\sqrt{2} \alpha _3 \beta_2+\sqrt{3} \alpha _4 \beta_1
\end{array}
\right)
\end{equation}

\begin{equation}
\nonumber\mathbf{5}_1\sim\left(
\begin{array}{c}
 \sqrt{2} \alpha _1 \beta _5-\sqrt{2} \alpha _2 \beta_4-\sqrt{2} \alpha _3 \beta_3+\sqrt{2} \alpha _4 \beta_2
   \\
 -\sqrt{2} \alpha _1 \beta_1-\sqrt{3} \alpha _3 \beta _4 -\sqrt{3} \alpha _4 \beta_3\\
 \sqrt{3} \alpha _1 \beta_2+\sqrt{2} \alpha _2 \beta_1+\sqrt{3} \alpha _3 \beta _5
   \\
 \sqrt{3} \alpha _2 \beta_2+\sqrt{2} \alpha _3 \beta_1+\sqrt{3} \alpha _4 \beta _5
   \\
 -\sqrt{3} \alpha _1 \beta _4-\sqrt{3} \alpha _2 \beta_3-\sqrt{2} \alpha _4 \beta_1
\end{array}
\right)
\end{equation}

\begin{equation}
\nonumber\mathbf{5}_2\sim\left(
\begin{array}{c}
 2 \alpha _1\beta _5+4 \alpha _2 \beta _4+4 \alpha _3 \beta_3 +2 \alpha _4 \beta _2\\
 4 \alpha _1 \beta _1+2 \sqrt{6} \alpha_2 \beta _5 \\
 -\sqrt{6} \alpha _1\beta _2+2 \alpha _2 \beta _1-\sqrt{6} \alpha _3 \beta _5 +2 \sqrt{6} \alpha _4 \beta_4\\
 2 \sqrt{6} \alpha _1 \beta_3-\sqrt{6} \alpha _2\beta _2+2 \alpha _3 \beta _1-\sqrt{6} \alpha _4 \beta _5 \\
 2 \sqrt{6} \alpha_3 \beta _2+4 \alpha _4 \beta _1
\end{array}
\right)
\end{equation}

\fbox{$\mathbf{5}\otimes\mathbf{5}=\mathbf{1}_S\oplus\mathbf{3}_A\oplus\mathbf{3}'_A\oplus\mathbf{4}_{S}\oplus\mathbf{4}_{A}\oplus\mathbf{5}_{S,1}\oplus\mathbf{5}_{S,2}$}

\begin{equation}
\nonumber\mathbf{1}_S\sim\alpha_1\beta_1+\alpha_2\beta_5+\alpha_3\beta_4+\alpha_4\beta_3+\alpha_5\beta_2
\end{equation}

\begin{equation}
\nonumber\mathbf{3}_A\sim\left(
\begin{array}{c}
 \alpha _2 \beta _5+2\alpha _3 \beta _4-2 \alpha _4 \beta _3-\alpha _5 \beta _2 \\
 -\sqrt{3} \alpha _1\beta _2+\sqrt{3} \alpha _2 \beta _1+\sqrt{2}\alpha _3 \beta _5-\sqrt{2} \alpha _5 \beta _3 \\
 \sqrt{3}\alpha _1 \beta _5+\sqrt{2} \alpha _2 \beta _4-\sqrt{2} \alpha _4\beta _2-\sqrt{3} \alpha _5 \beta _1
\end{array}
\right)
\end{equation}

\begin{equation}
\nonumber\mathbf{3}'_A\sim\left(
\begin{array}{c}
 2 \alpha _2 \beta _5-\alpha_3 \beta _4+\alpha _4 \beta _3-2 \alpha _5 \beta _2\\
 \sqrt{3} \alpha _1\beta _3-\sqrt{3} \alpha _3 \beta _1+\sqrt{2}\alpha _4 \beta _5-\sqrt{2} \alpha _5 \beta _4 \\
 -\sqrt{3}\alpha _1 \beta _4+\sqrt{2} \alpha _2 \beta _3-\sqrt{2} \alpha _3\beta _2+\sqrt{3} \alpha _4 \beta _1
\end{array}
\right)
\end{equation}

\begin{equation}
\nonumber\mathbf{4}_S\sim\left(
\begin{array}{c}
 3 \sqrt{2} \alpha _1 \beta _2+3 \sqrt{2} \alpha _2 \beta _1-\sqrt{3}\alpha_3\beta _5+4 \sqrt{3} \alpha _4 \beta _4-\sqrt{3}\alpha _5 \beta _3 \\
 3 \sqrt{2}\alpha _1 \beta _3+4 \sqrt{3} \alpha _2 \beta _2+3 \sqrt{2} \alpha _3 \beta _1-\sqrt{3}\alpha_4\beta _5-\sqrt{3} \alpha _5 \beta _4 \\
 3 \sqrt{2} \alpha _1 \beta _4-\sqrt{3}\alpha _2 \beta _3-\sqrt{3} \alpha _3 \beta _2+3 \sqrt{2} \alpha _4 \beta _1+4 \sqrt{3}\alpha _5\beta _5 \\
 3 \sqrt{2} \alpha_1\beta _5-\sqrt{3} \alpha _2 \beta _4+4 \sqrt{3}\alpha _3 \beta _3-\sqrt{3} \alpha _4 \beta _2+3 \sqrt{2} \alpha _5 \beta _1
\end{array}
\right)
\end{equation}

\begin{equation}
\nonumber\mathbf{4}_A\sim\left(
\begin{array}{c}
 \sqrt{2} \alpha _1 \beta _2-\sqrt{2} \alpha _2 \beta _1+\sqrt{3} \alpha _3 \beta _5-\sqrt{3}\alpha _5 \beta _3 \\
 -\sqrt{2} \alpha _1 \beta _3+\sqrt{2} \alpha _3 \beta _1+\sqrt{3} \alpha _4 \beta _5-\sqrt{3} \alpha_5 \beta _4 \\
 -\sqrt{2} \alpha _1 \beta _4-\sqrt{3} \alpha_2 \beta _3+\sqrt{3} \alpha _3 \beta _2+\sqrt{2} \alpha _4 \beta _1\\
 \sqrt{2} \alpha _1 \beta _5-\sqrt{3}\alpha _2 \beta _4+\sqrt{3} \alpha _4 \beta _2-\sqrt{2} \alpha _5 \beta _1
\end{array}
\right)
\end{equation}

\begin{equation}
\nonumber\mathbf{5}_{S,1}\sim\left(
\begin{array}{c}
 2 \alpha _1 \beta _1+\alpha _2 \beta _5-2 \alpha_3 \beta _4-2 \alpha _4 \beta _3+\alpha _5 \beta _2 \\
 \alpha _1 \beta _2+\alpha _2 \beta _1+\sqrt{6} \alpha _3 \beta _5+\sqrt{6} \alpha _5 \beta_3 \\
 -2 \alpha _1 \beta _3+\sqrt{6} \alpha _2 \beta _2-2 \alpha _3 \beta _1 \\
 -2 \alpha _1 \beta _4-2 \alpha _4 \beta _1+\sqrt{6} \alpha _5 \beta _5 \\
 \alpha _1 \beta _5+\sqrt{6} \alpha _2 \beta_4+\sqrt{6} \alpha _4 \beta _2+\alpha _5 \beta _1
\end{array}
\right)
\end{equation}

\begin{equation}
\nonumber\mathbf{5}_{S,2}\sim\left(
\begin{array}{c}
 2 \alpha _1 \beta _1-2 \alpha _2 \beta _5+\alpha _3 \beta _4+\alpha _4\beta _3-2 \alpha _5 \beta _2 \\
 -2 \alpha _1 \beta _2-2 \alpha _2 \beta _1+\sqrt{6} \alpha _4 \beta _4 \\
 \alpha _1 \beta _3+\alpha _3 \beta _1+\sqrt{6} \alpha _4 \beta _5+\sqrt{6} \alpha _5 \beta_4 \\
 \alpha _1 \beta _4+\sqrt{6} \alpha _2 \beta_3+\sqrt{6} \alpha _3 \beta _2+\alpha _4 \beta _1 \\
 -2 \alpha _1 \beta _5+\sqrt{6} \alpha _3 \beta _3-2 \alpha _5 \beta _1
 \end{array}
\right).
\end{equation}


\begin{thebibliography}{99}

\bibitem{Maki:1962mu}
  Z.~Maki, M.~Nakagawa and S.~Sakata,
  Prog.\ Theor.\ Phys.\  {\bf 28}, 870 (1962).
\bibitem{Pontecorvo:1967fh}
  B.~Pontecorvo,
  Sov.\ Phys.\ JETP {\bf 26}, 984 (1968)
  [Zh.\ Eksp.\ Teor.\ Fiz.\  {\bf 53}, 1717 (1967)].


\bibitem{Schwetz:2011zk}
  T.~Schwetz, M.~Tortola, J.~W.~F.~Valle,
[arXiv:1108.1376 [hep-ph]].

\bibitem{Schwetz:2011qt}
  T.~Schwetz, M.~Tortola and J.~W.~F.~Valle,
  New J.\ Phys.\  {\bf 13}, 063004 (2011)
  [arXiv:1103.0734 [hep-ph]];  M.~Maltoni and T.~Schwetz,
  arXiv:0812.3161 [hep-ph];
  T.~Schwetz, M.~A.~Tortola and J.~W.~F.~Valle,
  New J.\ Phys.\  {\bf 10}, 113011 (2008)
  [arXiv:0808.2016 [hep-ph]].

\bibitem{Fogli:2011qn}
  G.~L.~Fogli, E.~Lisi, A.~Marrone, A.~Palazzo and A.~M.~Rotunno,
  arXiv:1106.6028 [hep-ph].

\bibitem{Fogli:Indication}
G.~L.~Fogli, E.~Lisi, A.~Marrone, A.~Palazzo and A.~M.~Rotunno,
  arXiv:0809.2936 [hep-ph];
G.~L.~Fogli, E.~Lisi, A.~Marrone, A.~Palazzo and A.~M.~Rotunno,
  Phys.\ Rev.\ Lett.\  {\bf 101} (2008) 141801
  [arXiv:0806.2649 [hep-ph]].

\bibitem{GonzalezGarcia:2010er}
  M.~C.~Gonzalez-Garcia, M.~Maltoni and J.~Salvado,
  JHEP {\bf 1004}, 056 (2010)
  [arXiv:1001.4524 [hep-ph]].

\bibitem{Abe:2011sj}
  K.~Abe {\it et al.}  [T2K Collaboration],
  Phys.\ Rev.\ Lett.\  {\bf 107}, 041801 (2011)
  [arXiv:1106.2822 [hep-ex]].

\bibitem{minos}
MINOS Collaboration, http://www-numi.fnal.gov/pr\_plots/index.html.


\bibitem{double_chooz}
Talk given by H. de Kerret at the Sixth International Workshop on Low Energy
Neutrino Physics (LowNu11) at Seoul, Korea during November 9-12, 2011.


\bibitem{Ardellier:2006mn}
  F.~Ardellier {\it et al.}  [Double Chooz Collaboration],
  arXiv:hep-ex/0606025.

\bibitem{Wang:2006ca}
  Y.~f.~Wang,
  arXiv:hep-ex/0610024.


\bibitem{review}
  G.~Altarelli and F.~Feruglio,
  New J.\ Phys.\  {\bf 6}, 106 (2004)
  [arXiv:hep-ph/0405048];
  R.~N.~Mohapatra {\it et al.},
  Rept.\ Prog.\ Phys.\  {\bf 70}, 1757 (2007)
  [arXiv:hep-ph/0510213];
  A.~Strumia and F.~Vissani,
  arXiv:hep-ph/0606054.
  E.~Ma,
  arXiv:0705.0327 [hep-ph];
  G.~Altarelli,
  arXiv:0705.0860 [hep-ph];
  G.~Altarelli,
  arXiv:0711.0161 [hep-ph];
  F.~Feruglio, C.~Hagedorn, Y.~Lin and L.~Merlo,
  arXiv:0808.0812 [hep-ph];
  G.~Altarelli,
  arXiv:0905.2350 [hep-ph];
  G.~Altarelli,
  Nuovo Cim.\  C {\bf 32N5-6}, 91 (2009)
  [arXiv:0905.3265 [hep-ph]];
  S.~F.~King,
  AIP Conf.\ Proc.\  {\bf 1200}, 103 (2010)
  [arXiv:0909.2969 [hep-ph]];
  G.~Altarelli and F.~Feruglio,
  arXiv:1002.0211 [hep-ph];
  H.~Ishimori, T.~Kobayashi, H.~Ohki, H.~Okada, Y.~Shimizu and M.~Tanimoto,
  Prog.\ Theor.\ Phys.\ Suppl.\  {\bf 183}, 1 (2010)
  [arXiv:1003.3552 [hep-th]];
  M.~-C.~Chen, K.~T.~Mahanthappa,
   [arXiv:1012.1595 [hep-ph]].

\bibitem{A4}
  E.~Ma and G.~Rajasekaran,
  Phys.\ Rev.\  D {\bf 64}, 113012 (2001)
  [arXiv:hep-ph/0106291];
  K.~S.~Babu, E.~Ma and J.~W.~F.~Valle,
  Phys.\ Lett.\  B {\bf 552}, 207 (2003)
  [arXiv:hep-ph/0206292];
  E.~Ma,
  Phys.\ Rev.\  D {\bf 70}, 031901 (2004)
  [arXiv:hep-ph/0404199].
  G.~Altarelli and F.~Feruglio,
  Nucl.\ Phys.\  B {\bf 720}, 64 (2005)
  [arXiv:hep-ph/0504165];
  E.~Ma,
  Phys.\ Rev.\  D {\bf 72}, 037301 (2005)
  [arXiv:hep-ph/0505209];
  A.~Zee,
  Phys.\ Lett.\  B {\bf 630}, 58 (2005)
  [arXiv:hep-ph/0508278].
  G.~Altarelli and F.~Feruglio,
  Nucl.\ Phys.\  B {\bf 741}, 215 (2006)
  [arXiv:hep-ph/0512103].
  E.~Ma,
  Mod.\ Phys.\ Lett.\  A {\bf 21}, 2931 (2006)
  [arXiv:hep-ph/0607190];
  S.~F.~King and M.~Malinsky,
  Phys.\ Lett.\  B {\bf 645}, 351 (2007)
  [arXiv:hep-ph/0610250];
  S.~Morisi, M.~Picariello and E.~Torrente-Lujan,
  Phys.\ Rev.\  D {\bf 75}, 075015 (2007)
  [arXiv:hep-ph/0702034].
  M.~Honda and M.~Tanimoto,
  Prog.\ Theor.\ Phys.\  {\bf 119}, 583 (2008)
  [arXiv:0801.0181 [hep-ph]];
  G.~Altarelli, F.~Feruglio and C.~Hagedorn,
  JHEP {\bf 0803}, 052 (2008)
  [arXiv:0802.0090 [hep-ph]];
  P.~H.~Frampton and S.~Matsuzaki,
  arXiv:0806.4592 [hep-ph];
  F.~Feruglio, C.~Hagedorn, Y.~Lin and L.~Merlo,
  Nucl.\ Phys.\  B {\bf 809}, 218 (2009)
  [arXiv:0807.3160 [hep-ph]].
  P.~Ciafaloni, M.~Picariello, E.~Torrente-Lujan and A.~Urbano,
  Phys.\ Rev.\  D {\bf 79}, 116010 (2009)
  [arXiv:0901.2236 [hep-ph]];
  C.~Hagedorn, E.~Molinaro and S.~T.~Petcov,
  JHEP {\bf 0909}, 115 (2009)
  [arXiv:0908.0240 [hep-ph]];
  T.~J.~Burrows and S.~F.~King,
  Nucl.\ Phys.\  B {\bf 835}, 174 (2010)
  [arXiv:0909.1433 [hep-ph]];
  F.~Feruglio, C.~Hagedorn and L.~Merlo,
  JHEP {\bf 1003}, 084 (2010)
  [arXiv:0910.4058 [hep-ph]];
  J.~Berger and Y.~Grossman,
  JHEP {\bf 1002}, 071 (2010)
  [arXiv:0910.4392 [hep-ph]];
  F.~Feruglio, C.~Hagedorn, Y.~Lin and L.~Merlo,
  arXiv:0911.3874 [hep-ph];
  M.~Mitra,
  JHEP {\bf 1011}, 026 (2010)
  [arXiv:0912.5291 [hep-ph]];
T.~Fukuyama, H.~Sugiyama and K.~Tsumura,
  Phys.\ Rev.\  D {\bf 82}, 036004 (2010)
  [arXiv:1005.5338 [hep-ph]]; T.~Fukuyama, H.~Sugiyama and K.~Tsumura,
  Phys.\ Rev.\  D {\bf 83}, 056016 (2011)
  [arXiv:1012.4886 [hep-ph]];
 S.~Antusch, S.~F.~King, C.~Luhn and M.~Spinrath,
  Nucl.\ Phys.\  B {\bf 850}, 477 (2011)
  [arXiv:1103.5930 [hep-ph]];
  I.~K.~Cooper, S.~F.~King and C.~Luhn,
  Phys.\ Lett.\  B {\bf 690}, 396 (2010)
  [arXiv:1004.3243 [hep-ph]];
  A.~Kadosh and E.~Pallante,
  JHEP {\bf 1008}, 115 (2010)
  [arXiv:1004.0321 [hep-ph]];
S.~Antusch, S.~F.~King, M.~Spinrath,
  Phys.\ Rev.\  {\bf D83}, 013005 (2011).
  [arXiv:1005.0708 [hep-ph]];
  T.~J.~Burrows and S.~F.~King,
  Nucl.\ Phys.\  B {\bf 842}, 107 (2011)
  [arXiv:1007.2310 [hep-ph]];
  G.~-J.~Ding, D.~Meloni,
[arXiv:1108.2733 [hep-ph]].


\bibitem{tprime}
  A.~Aranda, C.~D.~Carone and R.~F.~Lebed,
  Phys.\ Lett.\  B {\bf 474}, 170 (2000)
  [arXiv:hep-ph/9910392].
  A.~Aranda, C.~D.~Carone and R.~F.~Lebed,
  Int.\ J.\ Mod.\ Phys.\  A {\bf 16S1C}, 896 (2001)
  [arXiv:hep-ph/0010144].
  A.~Aranda, C.~D.~Carone and R.~F.~Lebed,
  Phys.\ Rev.\  D {\bf 62}, 016009 (2000)
  [arXiv:hep-ph/0002044].
  F.~Feruglio, C.~Hagedorn, Y.~Lin and L.~Merlo,
  Nucl.\ Phys.\  B {\bf 775}, 120 (2007)
  [Erratum-ibid.\  {\bf 836}, 127 (2010)]
  [arXiv:hep-ph/0702194].
  M.~C.~Chen and K.~T.~Mahanthappa,
  Phys.\ Lett.\  B {\bf 652}, 34 (2007)
  [arXiv:0705.0714 [hep-ph]];
  A.~Aranda,
  Phys.\ Rev.\  D {\bf 76}, 111301 (2007)
  [arXiv:0707.3661 [hep-ph]];
  P.~H.~Frampton and T.~W.~Kephart,
  JHEP {\bf 0709}, 110 (2007)
  [arXiv:0706.1186 [hep-ph]];
  M.~C.~Chen and K.~T.~Mahanthappa,
  arXiv:0710.2118 [hep-ph];
  S.~Sen,
  Phys.\ Rev.\  D {\bf 76}, 115020 (2007)
  [arXiv:0710.2734 [hep-ph]];
  G.~J.~Ding,
  Phys.\ Rev.\  D {\bf 78}, 036011 (2008)
  [arXiv:0803.2278 [hep-ph]];
  P.~H.~Frampton, T.~W.~Kephart and S.~Matsuzaki,
  Phys.\ Rev.\  D {\bf 78}, 073004 (2008)
  [arXiv:0807.4713 [hep-ph]];
  M.~C.~Chen and K.~T.~Mahanthappa,
  Phys.\ Lett.\  B {\bf 681}, 444 (2009)
  [arXiv:0904.1721 [hep-ph]];
  M.~C.~Chen, K.~T.~Mahanthappa and F.~Yu,
  Phys.\ Rev.\  D {\bf 81}, 036004 (2010)
  [arXiv:0907.3963 [hep-ph]];
  L.~Merlo,
  arXiv:1004.2211 [hep-ph];
  M.~-C.~Chen, K.~T.~Mahanthappa,
  PoS {\bf ICHEP2010}, 407 (2010).
  [arXiv:1011.6364 [hep-ph]];
  M.~-C.~Chen, K.~T.~Mahanthappa,
  [arXiv:1107.3856 [hep-ph]];
  M.~-C.~Chen, K.~T.~Mahanthappa, A.~Meroni, S.~T.~Petcov,
    [arXiv:1109.0731 [hep-ph]].



\bibitem{delta}
  E.~Ma,
  Mod.\ Phys.\ Lett.\  A {\bf 21}, 1917 (2006)
  [arXiv:hep-ph/0607056].
  I.~de Medeiros Varzielas, S.~F.~King and G.~G.~Ross,
  Phys.\ Lett.\  B {\bf 648}, 201 (2007)
  [arXiv:hep-ph/0607045];
  C.~Luhn, S.~Nasri and P.~Ramond,
  J.\ Math.\ Phys.\  {\bf 48}, 073501 (2007)
  [arXiv:hep-th/0701188];
  E.~Ma,
  Phys.\ Lett.\  B {\bf 660}, 505 (2008)
  [arXiv:0709.0507 [hep-ph]];
  J.~A.~Escobar and C.~Luhn,
  J.\ Math.\ Phys.\  {\bf 50}, 013524 (2009)
  [arXiv:0809.0639 [hep-th]].
  S.~F.~King and C.~Luhn,
  JHEP {\bf 0910}, 093 (2009)
  [arXiv:0908.1897 [hep-ph]].



\bibitem{Z3xZ7}
  C.~Luhn, S.~Nasri and P.~Ramond,
  Phys.\ Lett.\  B {\bf 652}, 27 (2007)
  [arXiv:0706.2341 [hep-ph]];
  C.~Hagedorn, M.~A.~Schmidt and A.~Y.~Smirnov,
  Phys.\ Rev.\  D {\bf 79}, 036002 (2009)
  [arXiv:0811.2955 [hep-ph]].

\bibitem{Kajiyama:2007gx}
  Y.~Kajiyama, M.~Raidal and A.~Strumia,
  Phys.\ Rev.\  D {\bf 76}, 117301 (2007)
  [arXiv:0705.4559 [hep-ph]].



\bibitem{S4}
  E.~Ma,
  Phys.\ Lett.\  B {\bf 632}, 352 (2006)
  [arXiv:hep-ph/0508231];
  C.~Hagedorn, M.~Lindner and R.~N.~Mohapatra,
  JHEP {\bf 0606}, 042 (2006)
  [arXiv:hep-ph/0602244];
  Y.~Cai and H.~B.~Yu,
  Phys.\ Rev.\  D {\bf 74}, 115005 (2006)
  [arXiv:hep-ph/0608022];
  H.~Zhang,
  Phys.\ Lett.\  B {\bf 655}, 132 (2007)
  [arXiv:hep-ph/0612214];
  Y.~Koide,
  JHEP {\bf 0708}, 086 (2007)
  [arXiv:0705.2275 [hep-ph]];
  F.~Bazzocchi, L.~Merlo and S.~Morisi,
  Nucl.\ Phys.\  B {\bf 816}, 204 (2009)
  [arXiv:0901.2086 [hep-ph]];
  G.~-J.~Ding,
  Nucl.\ Phys.\  {\bf B827}, 82-111 (2010).
  [arXiv:0909.2210 [hep-ph]];
  Y.~Daikoku and H.~Okada,
  Phys.\ Rev.\  D {\bf 82}, 033007 (2010)
  [arXiv:0910.3370 [hep-ph]];
  C.~Hagedorn, S.~F.~King and C.~Luhn,
  JHEP {\bf 1006}, 048 (2010)
  [arXiv:1003.4249 [hep-ph]];
  R.~de Adelhart Toorop, F.~Bazzocchi and L.~Merlo,
  JHEP {\bf 1008}, 001 (2010)
  [arXiv:1003.4502 [hep-ph]];
  G.~J.~Ding,
  arXiv:1006.4800 [hep-ph];
  Y.~Daikoku and H.~Okada,
  arXiv:1008.0914 [hep-ph].




\bibitem{S3}
  P.~F.~Harrison and W.~G.~Scott,
  Phys.\ Lett.\  B {\bf 557}, 76 (2003)
  [arXiv:hep-ph/0302025];
  J.~Kubo, A.~Mondragon, M.~Mondragon and E.~Rodriguez-Jauregui,
  Prog.\ Theor.\ Phys.\  {\bf 109}, 795 (2003)
  [Erratum-ibid.\  {\bf 114}, 287 (2005)]
  [arXiv:hep-ph/0302196].
  T.~Kobayashi, J.~Kubo and H.~Terao,
  Phys.\ Lett.\  B {\bf 568}, 83 (2003)
  [arXiv:hep-ph/0303084];
  S.~L.~Chen, M.~Frigerio and E.~Ma,
  Phys.\ Rev.\  D {\bf 70}, 073008 (2004)
  [Erratum-ibid.\  D {\bf 70}, 079905 (2004)]
  [arXiv:hep-ph/0404084];
  F.~Caravaglios and S.~Morisi,
  arXiv:hep-ph/0503234;
  S.~Morisi and M.~Picariello,
  Int.\ J.\ Theor.\ Phys.\  {\bf 45}, 1267 (2006)
  [arXiv:hep-ph/0505113];
  Y.~Koide,
  Phys.\ Rev.\  D {\bf 73}, 057901 (2006)
  [arXiv:hep-ph/0509214];
  N.~Haba and K.~Yoshioka,
  Nucl.\ Phys.\  B {\bf 739}, 254 (2006)
  [arXiv:hep-ph/0511108];
  R.~N.~Mohapatra, S.~Nasri and H.~B.~Yu,
  Phys.\ Lett.\  B {\bf 639}, 318 (2006)
  [arXiv:hep-ph/0605020];
  Y.~Koide,
  Eur.\ Phys.\ J.\  C {\bf 50}, 809 (2007)
  [arXiv:hep-ph/0612058];
 A.~Mondragon, M.~Mondragon and E.~Peinado,
  Phys.\ Rev.\  D {\bf 76}, 076003 (2007)
  [arXiv:0706.0354 [hep-ph]];
  C.~Y.~Chen and L.~Wolfenstein,
  Phys.\ Rev.\  D {\bf 77}, 093009 (2008)
  [arXiv:0709.3767 [hep-ph]];
  M.~Mitra and S.~Choubey,
  arXiv:0806.3254 [hep-ph];
  D.~A.~Dicus, S.~F.~Ge and W.~W.~Repko,
  Phys.\ Rev.\  D {\bf 82}, 033005 (2010)
  [arXiv:1004.3266 [hep-ph]].

\bibitem{S3xA4}
  K.~S.~Babu and S.~Gabriel,
  arXiv:1006.0203 [hep-ph].

  \bibitem{PSL27}
  S.~F.~King and C.~Luhn,
  Nucl.\ Phys.\  B {\bf 820}, 269 (2009)
  [arXiv:0905.1686 [hep-ph]].
  S.~F.~King and C.~Luhn,
  Nucl.\ Phys.\  B {\bf 832}, 414 (2010)
  [arXiv:0912.1344 [hep-ph]].


\bibitem{Quat}
  P.~H.~Frampton and O.~C.~W.~Kong,
  Phys.\ Rev.\  D {\bf 53}, 2293 (1996)
  [arXiv:hep-ph/9511343];
  P.~H.~Frampton and O.~C.~W.~Kong,
  Phys.\ Rev.\ Lett.\  {\bf 75}, 781 (1995)
  [arXiv:hep-ph/9502395];
  P.~H.~Frampton and A.~Rasin,
  Phys.\ Lett.\  B {\bf 478}, 424 (2000)
  [arXiv:hep-ph/9910522];
  M.~Frigerio, S.~Kaneko, E.~Ma and M.~Tanimoto,
  Phys.\ Rev.\  D {\bf 71}, 011901 (2005)
  [arXiv:hep-ph/0409187];
  M.~Frigerio,
  arXiv:hep-ph/0505144;
  K.~S.~Babu and J.~Kubo,
  Phys.\ Rev.\  D {\bf 71}, 056006 (2005)
  [arXiv:hep-ph/0411226];
  M.~Frigerio and E.~Ma,
  Phys.\ Rev.\  D {\bf 76}, 096007 (2007)
  [arXiv:0708.0166 [hep-ph]]; A.~Aranda, C.~Bonilla, R.~Ramos and A.~D.~Rojas,
  Phys.\ Rev.\  D {\bf 84}, 016009 (2011)
  [arXiv:1105.6373 [hep-ph]].




\bibitem{D}
  C.~D.~Carone and R.~F.~Lebed,
  Phys.\ Rev.\  D {\bf 60}, 096002 (1999)
  [arXiv:hep-ph/9905275];
  W.~Grimus and L.~Lavoura,
  Phys.\ Lett.\  B {\bf 572}, 189 (2003)
  [arXiv:hep-ph/0305046];
W.~Grimus, A.~S.~Joshipura, S.~Kaneko, L.~Lavoura and M.~Tanimoto,
  JHEP {\bf 0407}, 078 (2004)
  [arXiv:hep-ph/0407112].
  E.~Ma,
  Fizika B {\bf 14}, 35 (2005)
  [arXiv:hep-ph/0409288];
  S.~L.~Chen and E.~Ma,
  Phys.\ Lett.\  B {\bf 620}, 151 (2005)
  [arXiv:hep-ph/0505064];
  C.~Hagedorn, M.~Lindner and F.~Plentinger,
  Phys.\ Rev.\  D {\bf 74}, 025007 (2006)
  [arXiv:hep-ph/0604265];
  Y.~Kajiyama, J.~Kubo and H.~Okada,
  Phys.\ Rev.\  D {\bf 75}, 033001 (2007)
  [arXiv:hep-ph/0610072];
  P.~Ko, T.~Kobayashi, J.~h.~Park and S.~Raby,
  Phys.\ Rev.\  D {\bf 76}, 035005 (2007)
  [Erratum-ibid.\  D {\bf 76}, 059901 (2007)]
  [arXiv:0704.2807 [hep-ph]];
  A.~Blum, C.~Hagedorn and M.~Lindner,
  Phys.\ Rev.\  D {\bf 77}, 076004 (2008)
  [arXiv:0709.3450 [hep-ph]];
  A.~Blum, C.~Hagedorn and A.~Hohenegger,
  JHEP {\bf 0803}, 070 (2008)
  [arXiv:0710.5061 [hep-ph]];
  A.~Adulpravitchai, A.~Blum and C.~Hagedorn,
  JHEP {\bf 0903}, 046 (2009)
  [arXiv:0812.3799 [hep-ph]];
  A.~Blum and C.~Hagedorn,
  Nucl.\ Phys.\  B {\bf 821}, 327 (2009)
  [arXiv:0902.4885 [hep-ph]];
  C.~Hagedorn and R.~Ziegler,
  arXiv:1007.1888 [hep-ph].

\bibitem{GR2}
  A.~Adulpravitchai, A.~Blum and W.~Rodejohann,
  New J.\ Phys.\  {\bf 11}, 063026 (2009)
  [arXiv:0903.0531 [hep-ph]].

\bibitem{T13}
  G.~-J.~Ding,
  Nucl.\ Phys.\  {\bf B853}, 635-662 (2011).
  [arXiv:1105.5879 [hep-ph]].

\bibitem{A5everettstuart}
  L.~L.~Everett and A.~J.~Stuart,
  Phys.\ Rev.\  D {\bf 79}, 085005 (2009)
  [arXiv:0812.1057 [hep-ph]]

\bibitem{A54fam}
  C.~S.~Chen, T.~W.~Kephart and T.~C.~Yuan,
  JHEP {\bf 1104}, 015 (2011)
  [arXiv:1011.3199 [hep-ph]].

\bibitem{Feruglio:2011qq}
  F.~Feruglio and A.~Paris,
  JHEP {\bf 1103}, 101 (2011)
  [arXiv:1101.0393 [hep-ph]].




  \bibitem{Everett:2010rd}
  L.~L.~Everett and A.~J.~Stuart,
  Phys.\ Lett.\  B {\bf 698}, 131 (2011)
  [arXiv:1011.4928 [hep-ph]].


\bibitem{TBmix} P.~F.~Harrison, D.~H.~Perkins and W.~G.~Scott, Phys.\ Lett.\  B {\bf 530}, 167
(2002), hep-ph/0202074; P.~F.~Harrison and W.~G.~Scott, Phys.\
Lett.\  B {\bf 535}, 163 (2002), hep-ph/0203209; Z.~Z.~Xing, Phys.\
Lett.\  B {\bf 533}, 85 (2002), hep-ph/0204049; X.~G.~He and A.~Zee,
Phys.\ Lett.\  B {\bf 560}, 87 (2003), hep-ph/0301092.


\bibitem{TB_quark}
P. Kaus and S. Meshkov, Mod Phys. Lett. A 3 (1988) 1251;
P. Kaus and S. Meshkov, Phys. Rev. D 42 (1990) 1863.

\bibitem{S4t13}
  C.~Hagedorn, S.~F.~King and C.~Luhn,
  JHEP {\bf 1006}, 048 (2010)
  [arXiv:1003.4249 [hep-ph]].
D.~Meloni,
  [arXiv:1107.0221 [hep-ph]];
  S.~Morisi, K.~M.~Patel and E.~Peinado,
  arXiv:1107.0696 [hep-ph].
  P.~S.~Bhupal Dev, R.~N.~Mohapatra and M.~Severson,
  arXiv:1107.2378 [hep-ph]; S.~F.~Ge, D.~A.~Dicus and W.~W.~Repko,
  arXiv:1108.0964 [hep-ph].


\bibitem{A4t13}
  Y.~Shimizu, M.~Tanimoto and A.~Watanabe,
  Prog.\ Theor.\ Phys.\  {\bf 126}, 81 (2011)
  [arXiv:1105.2929 [hep-ph]].
  E.~Ma and D.~Wegman,
  Phys.\ Rev.\ Lett.\  {\bf 107}, 061803 (2011)
  [arXiv:1106.4269 [hep-ph]].
  I.~de Medeiros Varzielas and L.~Merlo,
  JHEP {\bf 1102}, 062 (2011)
  [arXiv:1011.6662 [hep-ph]].

\bibitem{S3t13}
  S.~Zhou,
  arXiv:1106.4808 [hep-ph].


\bibitem{SU5t13} 
  S.~Antusch and V.~Maurer,
  arXiv:1107.3728 [hep-ph]; D.~Marzocca, S.~T.~Petcov, A.~Romanino and M.~Spinrath,
  arXiv:1108.0614 [hep-ph].

\bibitem{A4S4t13}
S.~F.~King and C.~Luhn,
  JHEP {\bf 1109}, 042 (2011)
  [arXiv:1107.5332 [hep-ph]];
  S.~Antusch, S.~F.~King, C.~Luhn and M.~Spinrath,
  arXiv:1108.4278 [hep-ph].




\bibitem{model.ind.t13}
  S.~F.~Ge, D.~A.~Dicus and W.~W.~Repko,
  Phys.\ Lett.\  B {\bf 702}, 220 (2011)
  [arXiv:1104.0602 [hep-ph]].
  H.~J.~He and F.~R.~Yin,
  Phys.\ Rev.\  D {\bf 84}, 033009 (2011)
  [arXiv:1104.2654 [hep-ph]].
  Z.~z.~Xing,
  arXiv:1106.3244 [hep-ph].
  X.~Chu, M.~Dhen and T.~Hambye,
  arXiv:1107.1589 [hep-ph].

\bibitem{Dt13}
  R.~d.~A.~Toorop, F.~Feruglio, C.~Hagedorn,
  Phys.\ Lett.\  {\bf B703}, 447-451 (2011).
  [arXiv:1107.3486 [hep-ph]].


  \bibitem{GRPrediction1}
  A.~Datta, F.~S.~Ling and P.~Ramond,
  Nucl.\ Phys.\  B {\bf 671}, 383 (2003)
  [arXiv:hep-ph/0306002].

\bibitem{Rodejohann:2008ir}
  W.~Rodejohann,
  Phys.\ Lett.\  B {\bf 671}, 267 (2009)
  [arXiv:0810.5239 [hep-ph]].



\bibitem{book:generator} H. Coxeter and W. Moser, Generators and Relations for Discrete Groups, Springer-Verlag,
Berlin (1957).

\bibitem{book:group} Wu-Ki Tung, Group Theory in Physics, World Scientific Publishing Company
(1985).

\bibitem{Lam:2008rs}
  C.~S.~Lam,
  Phys.\ Rev.\ Lett.\  {\bf 101}, 121602 (2008)
  [arXiv:0804.2622 [hep-ph]];  Phys.\ Rev.\  D {\bf 78}, 073015 (2008)
  [arXiv:0809.1185 [hep-ph]].


\bibitem{pdg} K.Nakamura et al. (Particle Data Group), J.P.G 37, 075021 (2010).


\bibitem{shirai}
  K.~Shirai,
J. \ Phys. \ Soc.\ Jpn.\ {\bf 61} 2735 (1992).

\bibitem{Barry:2010yk}
  J.~Barry, W.~Rodejohann,
  Nucl.\ Phys.\  {\bf B842}, 33-50 (2011).
  [arXiv:1007.5217 [hep-ph]].


\end{thebibliography}
\end{document}